\documentclass[11pt]{article}
\usepackage [utf8]{inputenc}
\usepackage{amsmath,amssymb,latexsym}
\usepackage{bbm}
\usepackage{hyperref}
\usepackage{enumerate}
\usepackage{nccmath}

\renewcommand{\thesection}{\arabic{section}}
\makeatletter \@addtoreset{equation}{section} \makeatother

\renewcommand{\theequation}{\thesection.\arabic{equation}}

\newcommand{\mytag}{\hfill\refstepcounter{equation}\textup{(\theequation)}}

\newtheorem{theorem}{Theorem}
\newtheorem{lemma}{Lemma}
\newtheorem{proposition}{Proposition}
\newtheorem{remark}{Remark}

\DeclareMathOperator{\tr}{tr}
\DeclareMathOperator{\End}{End}
\DeclareMathOperator{\diag}{diag}

\title{On the correlation functions of the characteristic polynomials of the sparse hermitian random matrices}
\author{Ie.\ Afanasiev\\ V. N. Karazin Kharkiv University, Kharkiv, Ukraine \\
Institute for Low Temperature Physics Ukr. Ac. Sci., Kharkiv, Ukraine\\ 
E-mail: afanasev\underline{ }e@mail.ru}
\date{ 
}

\begin{document}
\sloppy
%
%
%
%
%

\maketitle

\begin{abstract}
We consider asymptotics of the correlation functions of characteristic polynomials corresponding to random weighted $G(n, \frac{p}{n})$ Erd{\H o}s -- R\'enyi graphs with Gaussian weights in the case of finite $p$ and also when $p \to \infty$. It is shown that for finite $p$ the second correlation function demonstrates a kind of transition: when $p < 2$ it factorizes in the limit $n \to \infty$, while for $p > 2$ there appears an interval $(-\lambda_*(p), \lambda_*(p))$ such that for $\lambda_0 \in (-\lambda_*(p), \lambda_*(p))$ the second correlation function behaves like that for GUE, while for $\lambda_0$ outside the interval the second correlation function is still factorized. For $p \to \infty$ there is also a threshold in the behavior of the second correlation function near $\lambda_0 = \pm 2$: for $p \ll n^{2/3}$ the second correlation function factorizes, whereas for $p \gg n^{2/3}$ it behaves like that for GUE. For any rate of $p \to \infty$ the asymptotics of correlation functions of any even order for $\lambda_0 \in (-2, 2)$ coincide with that for GUE.
\end{abstract}
\section{Introduction and main results}

Consider an ensemble of hermitian $n \times n$ random matrices of the form
\begin{equation}
\label{ensemble M_n}
M_n = (d_{jk} w_{jk})_{j,k = 1}^{n}
\end{equation}
where
\begin{equation}
\begin{gathered}
\label{d_jk and w_jk}
d_{jk} = p^{-1/2} 
\begin{cases}
1\ \text{with probability}\, \dfrac{p}{n};\\
0\ \text{with probability}\, 1 - \dfrac{p}{n};
\end{cases} \\
w_{jk} = w_{jk}^{(1)} + iw_{jk}^{(2)},\, j \ne k
\end{gathered}
\end{equation}
and $\{w_{jk}^{(1)}, w_{jk}^{(2)}, w_{ll} : 1 \le j < k \le n, 1 \le l \le n\}$ are i.i.d.\ random variables with zero mean such that
\begin{equation}
\label{w_jk 1,2}
2\mathbf{E} \{|w_{jk}^{(1)}|^2\} = 2\mathbf{E} \{|w_{jk}^{(2)}|^2\} = \mathbf{E} \{|w_{ll}|^2\} = 1,\quad  j \ne k.
\end{equation}
Here and everywhere below $\mathbf{E}$ denotes the expectation with respect to all random variables. $\{d_{jk} : j \le k\}$ are also independent of each other and of $w_{jk}^{(1)}, w_{jk}^{(2)}, w_{ll}$.

These matrices are known as ``weighted'' adjacency matrices of random Erd{\H o}s -- R\'enyi $G(n, \frac{p}{n})$ graphs, with $\{d_{jk}\}$ corresponding to the standard adjacency matrix and $\{w_{jk}\}$ --- the set of independent weights, which we take to be Gaussian. These matrices are widely discussed in the last few years since they demonstrate a kind of interpolation between a ``sparse'' matrix with finite $p$, when there is only a finite number of nonzero elements in each line, and the matrix with $p = n$ coinciding with Gaussian Unitary Ensemble (GUE). The results on the convergence of normalized eigenvalue counting measure 
\begin{equation*}
N_n(\Delta) = \#\{\lambda_j^{(n)} \in \Delta,\, j = 1, \ldots, n\}/n, \quad N_n(\mathbb{R}) = 1
\end{equation*}
of these matrices in the case of finite $p$ were obtain in \cite{RB:88,RD:90} on the physical level of rigour, then in \cite{BG} for $w_{jk} = 1$ and in \cite{KSV} for arbitrary $\{w_{jk}\}$ independent on $\{d_{jk}\}$ and having 4 moments.

It was also shown that for $p \to \infty$ the limiting eigenvalue distribution coincides with GUE
\begin{equation}\label{eq:semicircle}
\lim\limits_{n \to \infty} N_n(\Delta) = \int\limits_\Delta \rho_{sc}(\lambda)d\lambda, \qquad \rho_{sc}(\lambda) = \frac{1}{2\pi}\sqrt{4 - \lambda^2}\cdot\mathbbm{1}_{[-2, 2]},
\end{equation}
while for finite $p$ the limiting measure is a solution of a rather complicated nonlinear integral equation which is difficult for the analysis. It is known that the support of the limiting measure (spectrum) is the whole real line. 
Central Limit Theorem for linear eigenvalue statistics was proven in \cite{Sh-Ti:10} for finite $p$ and in \cite{Sh-Ti:12} for $p \to \infty$. For the local regime it was conjectured the existence of the critical value $p_c > 1$ (see \cite{Ev-Ec:92}) such that for $p > p_c$ the eigenvalues are strongly correlated and are characterized by GUE matrix statistics, for $p < p_c$ the eigenvalues are uncorrelated and follow Poison statistics. The conjecture was confirmed by numerical calculations \cite{Ku:08} and by supersymmetry approach (SUSY) \cite{Mi-Fy:91,Mi-Fy:91(2)} on the physical lever of rigour. Notice, that the results of the present paper confirm the existence of similar threshold for the second correlation function of characteristic polynomials. Rigorous results for the local eigenvalue statistics were obtained recently in \cite{Er-Kn-Ya-Yi:12,Hu-La-Ya:15}. First for $p \gg n^{2/3}$ and then for $p \gg n^\varepsilon$ with any $\varepsilon > 0$ it was shown that the spectral correlation functions of sparse hermitian random matrices in the bulk of the spectrum converge in the weak sense to that of GUE. For the edge of the spectrum, it was proved in \cite{Kh:14} that for $p \gg n^{2/3}$ the limiting probability $P\{\max\limits_j\lambda_j^{(n)} > 2 + x/n^{2/3}\}$ admits a certain universal upper bound, whereas the result of \cite{Kh:15} implies that for $p \ll n^{1/5}$ the limiting probability $P\{\max\limits_j\lambda_j^{(n)} > 2 + x/p\}$ is zero. Note that more advanced results for the edge eigenvalue statistics were obtained in \cite{So:09} for so-called random $d$-regular graphs. It was shown that if $3 \le d \ll n^{2/3}$, and $w_{jk} = \pm 1$ then the scaled largest eigenvalue or \eqref{ensemble M_n} converges in distribution to the Tracy--Widom law.

The correlation functions of characteristic polynomials formally do not characterize the local eigenvalue statistics. However, from the SUSY point of view, their analysis is similar to that for spectral correlation functions. In combination with the fact that the analysis for correlation functions of characteristic polynomials usually is simpler than that for spectral correlation functions, it causes that such an analysis is often the first step in studies of local regimes.

The moments of the characteristic polynomials were studied for a lot of random matrix ensembles: for Gaussian Orthogonal Ensemble in \cite{Fy-Ke:03}, for Circular Unitary Ensemble in \cite{Ke-Sn:00,Fy-Kh:07}, for $\beta$-models with $\beta = 2$ in \cite{Br-Hi:00,St-Fy:03} and with $\beta = 1,\,2,\,4$ in \cite{Br-Hi:01,Me-No:01}. G\"otze and K\"osters in \cite{Go-Ko:08} have studied the second order correlation function of the characteristic polynomials of the Wigner matrix with arbitrary distributed entries, possessing the forth moments, by the method of generation functions. The result was generalized soon on the correlation functions of any even order by T.\ Shcherbina in \cite{Sh:11} where it was proposed the method which allowed to apply SUSY technique (or the Grassmann integration technique) to study the correlation functions of characteristic polynomials of random matrices with non Gaussian entries. The proposed method appeared to be rather powerful and since that was successfully applied to study characteristic polynomials of sample covariance matrices (see \cite{Sh:13}) and band matrices (\cite{Sh:14,Sh:15}).

In the present paper we apply the method of \cite{Sh:11} to study characteristic polynomials of sparse matrices. 
To be more precise, let us introduce our main definitions. The mixed moments or the correlation functions of the characteristic polynomials have the form
\begin{equation}
\label{F definition}
F_{2m}(\Lambda) = \mathbf{E}\bigg\{\prod\limits_{j = 1}^{2m} \det(M_n - \lambda_j)\bigg\},
\end{equation}
where $\Lambda = \operatorname{diag}\{\lambda_1, \ldots, \lambda_{2m}\}$ are real or complex parameters which may depend on $n$.

We are interested in the asymptotic behavior of \eqref{F definition} for matrices \eqref{ensemble M_n}, as $n \to \infty$, for
\begin{equation*}
\lambda_j = \lambda_0 + \frac{x_j}{n}, \quad j = \overline{1, 2m},
\end{equation*}
where $\lambda_0$, $\{x_j\}_{j = 1}^{2m}$ are real numbers and notation $j = \overline{1, 2m}$ means that $j$ varies from 1 to $2m$.

Set also
\begin{equation}
\label{D_and_b_2_definition}
D_{2m}(\Lambda) = \frac{F_{2m}(\Lambda)}{\Big(\prod\limits_{j = 1}^{2m}F_{2m}(\lambda_j I)\Big)^{1/2m}}, \qquad \lambda_*(p) = \left\{ \begin{array}{cc}
\sqrt{4 - 8/p}, & \text{if } p > 2;\\
0, & \text{if } p \le 2.
\end{array}\right.
\end{equation}

\begin{theorem}
\label{th_1}
Let an ensemble of sparse random matrices be defined by \eqref{ensemble M_n}-\eqref{w_jk 1,2} for finite~$p$ and let $w_{jk}^{(1)},\, w_{jk}^{(2)}$ be Gaussian random variables. Then the correlation function of two characteristic polynomials \eqref{F definition} for $m = 1$ satisfies the asymptotic relations
\begin{enumerate}[(i)]
\item for $\lambda_0 \in (-\lambda_*(p), \lambda_*(p))$
\begin{equation*}
\lim_{n \to \infty} D_2\left( \Lambda \right) = \frac{\sin((x_1-x_2)\sqrt{\lambda_*(p)^2-\lambda_0^2}/2)}{(x_1-x_2)\sqrt{\lambda_*(p)^2-\lambda_0^2}/2};
\end{equation*}

\item for $\lambda_0 \notin (-\lambda_*(p), \lambda_*(p))$
\begin{equation*}
\lim_{n \to \infty} D_2\left( \Lambda \right) = 1,
\end{equation*}

\end{enumerate}
where $D_2$ and $\lambda_*(p)$ are defined in \eqref{D_and_b_2_definition}.
\end{theorem}

\emph{Remarks}
\begin{enumerate}

\item The theorem shows that the second order correlation function has a threshold $p = 2$, i.e.\ if $p > 2$ there are two types of the asymptotic behavior --- cases (i) and (ii), if $p \le 2$ there is only one type of the asymptotic behavior --- case (ii).

\item If we let $\lambda_0$ depend on $n$, the asymptotic regimes (i) and (ii) are fully agreed.

\item Note that $\lambda_*(p) \to 2$, as $p \to \infty$, and since the limiting spectrum is always $[-2, 2]$ (see \eqref{eq:semicircle}), therefore for $p \to \infty$ one expects GUE behavior for all $\lambda_0 \in (-2, 2)$ (i.e. for all $\lambda_0$ in the interior of the limiting spectrum, cf \eqref{eq:semicircle}); we confirm this in Theorem \ref{th_3}.
\end{enumerate}

\begin{theorem}\label{th_3}
Let an ensemble of diluted random matrices be defined by \eqref{ensemble M_n}-\eqref{w_jk 1,2}, $p \to \infty$ and let $w_{jk}^{(1)},\, w_{jk}^{(2)}$ be Gaussian random variables. Then the correlation function of characteristic polynomials \eqref{F definition} for $\lambda_0 \in (-2, 2)$ satisfies the asymptotic relations
\begin{equation*}
\lim\limits_{n \to \infty} D_{2m}(\Lambda) = \frac{\hat{S}_{2m}(X)}{\hat{S}_{2m}(I)},
\end{equation*}
with $X = \diag\{x_1, \ldots, x_{2m}\}$ and
\begin{equation} \label{eq:hat S_2m definition}
\hat{S}_{2m}(X) = \frac{\det \left\{ \dfrac{\sin(\pi\rho_{sc}(\lambda_0)(x_j - x_{m + k}))}{\pi\rho_{sc}(\lambda_0)(x_j - x_{m + k})}\right\}_{j,k = 1}^m}{\Delta(x_1, \ldots, x_m)\Delta(x_{m + 1}, \ldots, x_{2m})}
\end{equation}
where $\Delta(y_1, \ldots, y_m)$ is the Vandermonde determinant of $y_1, \ldots, y_m$.
\end{theorem}

Notice that $\hat{S}_{2m}(I)$ is well defined because the difference of the rows $j_1$ and $j_2$ in the determinant in \eqref{eq:hat S_2m definition} is of order $O(x_{j_1} - x_{j_2})$, as $x_{j_1} \to x_{j_2}$. The same is true for columns.

To formulate our last result we introduce the Airy kernel
\begin{equation}
\label{eq:Airy kernel}
\mathbb{A}(x, y) = \frac{Ai(x)Ai'(y) - Ai'(x)Ai(y)}{x - y},
\end{equation}
where $Ai(x)$ is the Airy function.

\begin{theorem}
\label{th_2}
Let an ensemble of diluted random matrices be defined by \eqref{ensemble M_n}-\eqref{w_jk 1,2}, $p \to \infty$, and let $w_{jk}^{(1)},\, w_{jk}^{(2)}$ be Gaussian random variables. Then the correlation function of two characteristic polynomials \eqref{F definition} for $m = 1$ satisfies the asymptotic relations
\begin{itemize}

\item [(i)] If $\frac{n^{2/3}}{p} \to \infty$
$$
\lim_{n \to \infty} D_2(2I + n^{-2/3}X) = 1;
$$

\item [(ii)] If $\frac{n^{2/3}}{p} \to c$
\begin{equation*}
\lim_{n \to \infty} D_2(2I + n^{-2/3}X) = \frac{\mathbb{A}(x_1 + 2c, x_2 + 2c)}{\sqrt{\mathbb{A}(x_1 + 2c, x_1 + 2c) \mathbb{A}(x_2 + 2c, x_2 + 2c)}},
\end{equation*}
\end{itemize}
where $D_2$ is defined in \eqref{D_and_b_2_definition} and $\mathbb{A}$ is defined in \eqref{eq:Airy kernel}. For $\lambda_0 = -2$ similar assertions are also valid.
\end{theorem}

\emph{Remarks}

\begin{enumerate}
\item Notice, that the case $p \gg n^{2/3}$ corresponds to the case $c = 0$ in (ii).

\item The results of Theorem \ref{th_2} are in a good agreement with the results of \cite{Kh:14,Kh:15} in the sense that the asymptotic behavior changes when $p$ crosses the rate $n^{2/3}$. However, in \cite{Kh:15} it is argued that in the case $p \ll n^{2/3}$ the appropriate scale is $p^{-1}$ instead of $n^{-2/3}$. We postpone the study of $F_2$ with the scaling $p^{-1}$, as well as the related study of $F_2$ near $\lambda_*(p)$ for finite $p$, to subsequent publications.
\end{enumerate}

The paper is organized as follows. In Section \ref{sec:int_repres} we obtain a convenient integral representation for $F_{2m}$ using integration over the Grassmann variables and Harish-Chandra/Itzykson--Zuber formula for integrals over the unitary group. Sections \ref{sec:th1}, \ref{sec:th3} and \ref{sec:th2} deal with the proof of the Theorems \ref{th_1}, \ref{th_3} and \ref{th_2} respectively. The proof is based on the steepest descent method applied to the integral representation.

Notice also that everywhere below we denote by $C$ various $n$-independent constants, which can be different in different formulas.

\section{Integral representation}\label{sec:int_repres}

To formulate the result of the section, the following notations are introduced
\begin{itemize}
\item $\Delta(\diag\{y_j\}_{j = 1}^{k}) = \Delta(\{y_j\}_{j = 1}^{k})$ is the Vandermonde determinant of $\{y_j\}_{j = 1}^{k}$; \mytag\label{eq:Vandermonde definition}
\item $\End V$ is the set of linear operators on a linear space $V$;
\item $I_{n, k} = \{\alpha \in \mathbb{Z}^k | 1 \leq \alpha_1 < \ldots < \alpha_k \leq n\}$. \mytag\label{eq:I_nk definition}

The lexicographical order on $I_{n, k}$ is denoted by $\prec$;
\item $\mathcal{H}_{2m, l}$ is the space of self-adjoint operators in $\operatorname{End} \Lambda^l \mathbb{C}^{2m}$ (see \cite[Chapter 8.4]{Vi:03} for definition of $\Lambda^q V$);
\item $dB = dP_l(B) = \prod\limits_{\alpha \in I_{2m, l}} dB_{\alpha\alpha} \prod\limits_{\alpha \prec \beta} d\Re B_{\alpha\beta} d\Im B_{\alpha\beta}$ \mytag\label{eq:measure on H_2m,l}

is a measure on $\mathcal{H}_{2m, l}$. $B_{\alpha\beta}$ denotes the corresponding entry of the matrix of $B$ in some basis. It is easy to see that $dP_l(UBU^*) = dP_l(B)$ for any unitary matrix $U$, so the definition is correct.
\item $H_m = \prod\limits_{l = 2}^{2m} \mathcal{H}_{2m, l}$; \mytag\label{eq:H_m}
\item Set also
\begin{equation}
\label{eq:A_2m definition}
A_{2m}(G_1, G) = \sum\limits_{\substack{k_1 + 2k_2 + \ldots + 2mk_{2m} = 2m \\ k_j \in \mathbb{Z}_+}}(2m)!\prod\limits_{q = 1}^{2m} \frac{1}{(q!)^{k_q}k_q!} \bigwedge\limits_{s = 1}^{2m} (b_{s}G_{s} - \tilde{b}_sI)^{\wedge k_{s}}
\end{equation}
where $\{b_s\}_{s = 1}^\infty$, $\{\tilde{b}_s\}_{s = 1}^\infty$ are the sequences of certain $n, p$-dependent numbers and
\begin{align*}
G = (G_2, &\ldots, G_{2m}), \quad G_l \in \End \Lambda^l \mathbb{C}^{2m}, \quad l = \overline{1,2m}.
\end{align*}
Exterior product $A \wedge B$ of operators is defined in Section \ref{sec:grassmanns}. Since $\dim \Lambda^{2m} \mathbb{C}^{2m} = 1$, the space $\operatorname{End} \Lambda^{2m} \mathbb{C}^{2m}$ may be identified with the $\mathbb{C}$. In \eqref{eq:A_2m definition} $\bigwedge\limits_{s = 1}^{2m} (b_{s}G_{s} - \tilde{b}_sI)^{\wedge k_{s}}$ is understood as $\Big\{\bigwedge\limits_{s = 1}^{2m} (b_{s}G_{s} - \tilde{b}_sI)^{\wedge k_{s}}\Big\}_{1\ldots n;1\ldots n}$.
\item $C_n^{(2m)}(X) = \pi^m \left( \frac{1}{2} \right)^{\frac{1}{2}(2^{2m} - 1)} \big( \frac{n}{\pi} \big)^{\frac{1}{2}\left(\binom{4m}{2m} - 1\right)} \exp\Big\{\frac{1}{2n}\sum\limits_{j = 1}^{2m} x_j^2\Big\}$. \mytag\label{C_n^(2m)_definition}
\end{itemize}
We prove the following integral representation for the correlation function $F_{2m}$.

\begin{proposition}\label{pr1(integral representation)}
Let $M_n$ be a random matrix of the form \eqref{ensemble M_n}--\eqref{w_jk 1,2}, where $w_{jk}^{(1)}$, $w_{jk}^{(2)}$, $j < k$, $w_{ll}$ have Gaussian distribution. Then the correlation function \eqref{F definition} admits the representation
\begin{multline}
\label{final_integral_representation_m>1}
F_{2m}(\Lambda) = C_n^{(2m)}(X) \frac{i^{2m^2 - m}\exp\Big\{\lambda_0\sum\limits_{j = 1}^{2m} x_j\Big\}}{\Delta(X)} \\
\times \int\limits_{H_m}\int\limits_{\mathbb{R}^{2m}} \Delta(T) \exp\bigg\{-i\sum\limits_{j=1}^{2m} x_jt_j\bigg\} e^{nf_{2m}(T, R)} dT dR,
\end{multline}
where $R = (R_2, \ldots, R_{2m})$, $R_l \in \mathcal{H}_{2m, l}$, $dR = \prod\limits_{j = 2}^{2m} dR_j$, $T = \diag\{t_j\}_{j = 1}^{2m}$, $dT = \prod\limits_{j = 1}^{2m} dt_j$,
\begin{equation}
\label{f_2m_definition}
f_{2m}(T, R) = \log A_{2m}(T, R) - \frac{1}{2}\bigg(\sum\limits_{j=1}^{2m} (t_j+i\lambda_0)^2 + \sum\limits_{l = 2}^{2m} \operatorname{tr} R_l^2\bigg)
\end{equation}
and all other notation is defined at the beginning of the section.
\end{proposition}

\begin{remark}\label{rem1}
In the special case $m = 1$, the representation \eqref{final_integral_representation_m>1} simplifies to
\begin{equation}
\label{final_integral_representation}
F_2(\Lambda) = C_n(X) \frac{ie^{\lambda_0(x_1 + x_2)}}{x_1 - x_2}\int\limits_{\mathbb{R}^3} (t_1 - t_2) \exp\bigg\{-i\sum\limits_{j=1}^2 x_jt_j\bigg\} e^{nf(T, s)} dT ds,
\end{equation}
where $C_n(X) = C_n^{(2)}(X)$ and
\begin{equation}
\label{f_definition}
f(T, s) = \log(b_2 s - t_1 t_2) - \frac{1}{2}\bigg(\sum\limits_{j=1}^2 (t_j+i\lambda_0)^2 + s^2\bigg).
\end{equation}
\end{remark}

The proof of Proposition \ref{pr1(integral representation)} is based on the method of integration over the Grassmann variables, the required properties of which are reviewed in Section \ref{sec:grassmanns}. The proof of the proposition is given in Section \ref{sec:integral representation m=1}.

\subsection{Proof of Proposition \ref{pr1(integral representation)}}\label{sec:integral representation m=1}

Let us transform $F_{2m}(\Lambda)$, using \eqref{gauss_int}
\begin{multline*}
F_{2m}(\Lambda) = \mathbf{E} \bigg\{ \prod\limits_{j = 1}^{2m} \det(M - \lambda_j) \bigg\} = \mathbf{E} \bigg\{ \int\exp\bigg\{-\sum\limits_{l=1}^{2m} \bigg( \sum\limits_{1 \leq j, k \leq n} d_{jk}w_{jk}\overline{\psi}_{jl}\psi_{kl} \\
- \sum\limits_{j=1}^n \lambda_l \overline{\psi}_{jl}\psi_{jl} \bigg)\bigg\} \prod\limits_{l = 1}^{2m} \prod\limits_{j = 1}^n d\overline{\psi}_{jl}d\psi_{jl} \bigg\}
\end{multline*}

Averaging first with respect to $\{w_{jk}\}$, we obtain
\begin{multline*}
F_{2m}(\Lambda) = \mathbf{E}\bigg\{ \int \exp\bigg\{ \sum\limits_{j < k} d_{jk}^2\chi_{jk}\chi_{kj} 
+ \sum\limits_{j=1}^n \bigg(\frac{1}{2}d_{jj}^2\chi_{jj}^2 + \sum\limits_{l = 1}^{2m} \lambda_l\overline{\psi}_{jl}\psi_{jl}\bigg)\bigg\} \prod\limits_{l = 1}^{2m} \prod\limits_{j = 1}^n d\overline{\psi}_{jl}d\psi_{jl} \bigg\},
\end{multline*}
where, in order to simplify formulas below we denote
\begin{equation*}
\chi_{jk} = \sum\limits_{l = 1}^{2m} \overline{\psi}_{jl}\psi_{kl}.
\end{equation*}

Since evidently $(\chi_{jk}\chi_{kj})^{2m + 1} = 0$, we get
\begin{equation*}
\mathbf{E}\left\{ e^{d_{jk}^2 \chi_{jk}\chi_{kj}} \right\} = \mathbf{E}\bigg\{ 1 + \sum\limits_{l = 1}^{2m} \frac{1}{l!} d_{jk}^{2l} (\chi_{jk}\chi_{kj})^l \bigg\} = 1 + \sum\limits_{l = 1}^{2m} \frac{1}{l!} \cdot \frac{1}{p^{l - 1} n} (\chi_{jk}\chi_{kj})^l, \quad  j \leq k.
\end{equation*}

Define the numbers $\{a_l\}_{l=1}^{2m}$ by the identity
\begin{equation*}
\exp\bigg\{\sum\limits_{l = 1}^{2m} a_l(\chi_{jk}\chi_{kj})^l\bigg\} = 1 + \sum\limits_{l = 1}^{2m} \frac{1}{l!} \cdot \frac{1}{p^{l - 1} n} (\chi_{jk}\chi_{kj})^l.
\end{equation*}
Observe that
$$
a_1 = \frac{1}{n},\quad a_2 = \frac{n - p}{2pn^2}
$$
and that
\begin{equation}\label{eq:a_l asympt}
\begin{gathered}
a_l \sim \frac{C}{p^{l - 1}n},\, n \to \infty, \\
a_l = 0,\, l > 1,\, \text{if } p = n.
\end{gathered}
\end{equation}
Then $F_{2m}(\Lambda)$ can be represented as
\begin{multline}
\label{after_averaging_m>1}
F_{2m}(\Lambda) = \int \exp\bigg\{ \sum\limits_{j < k} \sum\limits_{l = 1}^{2m} a_l(\chi_{jk}\chi_{kj})^l + \sum\limits_{j=1}^n \bigg(\sum\limits_{l = 1}^m \frac{1}{2^l}a_l\chi_{jj}^{2l} + \sum\limits_{l = 1}^{2m} \lambda_l\overline{\psi}_{jl}\psi_{jl}\bigg) \bigg\} \\
\prod\limits_{l = 1}^{2m} \prod\limits_{j = 1}^n d\overline{\psi}_{jl}d\psi_{jl}.
\end{multline}

To facilitate the reading, the remaining steps are first explained in the simpler case $m = 1$ and only then in the general case.

\subsubsection{Case $m = 1$}

Let us transform the exponent of \eqref{after_averaging_m>1}.
\begin{align*}
\chi_{jk}\chi_{kj} &= \sum\limits_{\alpha, \beta \in I_{2,1}} \overline{\psi}_{j\alpha_1}\psi_{k\alpha_1}\overline{\psi}_{k\beta_1}\psi_{j\beta_1} = -\sum\limits_{\alpha, \beta \in I_{2,1}} \overline{\psi}_{j\alpha_1}\psi_{j\beta_1}\overline{\psi}_{k\beta_1}\psi_{k\alpha_1}, \\
(\chi_{jk}\chi_{kj})^2 &= 4\prod\limits_{l = 1}^2 \overline{\psi}_{jl}\psi_{kl}\overline{\psi}_{kl}\psi_{jl} = 4\prod\limits_{l = 1}^2 \overline{\psi}_{jl}\psi_{jl}\overline{\psi}_{kl}\psi_{kl} = 4\prod\limits_{l = 1}^2 \overline{\psi}_{jl}\psi_{jl} \prod\limits_{q = 1}^2 \overline{\psi}_{kq}\psi_{kq},
\end{align*}
where $I_{2, 1}$ is defined in \eqref{eq:I_nk definition}. Since $\psi_{jl}^2 = \overline{\psi}\vphantom{\psi}_{jl}^2 = 0$, we have
\begin{equation*}
\begin{gathered}
\sum\limits_{j < k} \chi_{jk}\chi_{kj} + \sum\limits_{j=1}^n \frac{1}{2}\chi_{jj}^2 = -\sum\limits_{l = 1}^2 \frac{1}{2}\bigg( \sum\limits_{j = 1}^n \overline{\psi}_{jl}\psi_{jl} \bigg)^2 - \bigg(\sum\limits_{j}\overline{\psi}_{j1}\psi_{j2} \bigg)\bigg(\sum\limits_{j}\overline{\psi}_{j2}\psi_{j1} \bigg). \\
\sum\limits_{j < k} (\chi_{jk}\chi_{kj})^2 = 2\bigg( \sum\limits_{j = 1}^n \prod\limits_{l = 1}^2 \overline{\psi}_{jl}\psi_{jl}\bigg)^2.
\end{gathered}
\end{equation*}

Hubbard-Stratonovich transformation \eqref{sqrt_of_exponent} applied to \eqref{after_averaging_m>1} yields
\begin{align*}
\exp\bigg\{ a_1\bigg(\sum\limits_{j < k} \chi_{jk}\chi_{kj} + \sum\limits_{j=1}^n \frac{1}{2}\chi_{jj}^2\bigg) \bigg\}& = \frac{n^2}{2\pi^2}\int\limits_{\mathcal{H}_2} \exp\bigg\{ -\frac{n}{2}\bigg(\sum\limits_{j = 1}^2 t_j^2 + 2(u^2 + v^2)\bigg) \bigg\}\\
\prod\limits_{j = 1}^n \exp\bigg\{ i\sum\limits_{l = 1}^2 t_l \overline{\psi}_{jl}&\psi_{jl} + i(u - iv)\overline{\psi}_{j2}\psi_{j1} + i(u + iv)\overline{\psi}_{j1}\psi_{j2} \bigg\} dQ;
\end{align*}
\begin{align*}
\exp\bigg\{ a_2\sum\limits_{j < k} (\chi_{jk}\chi_{kj})^2 \bigg\} = \sqrt{\frac{n}{2\pi}}\int\limits_{\mathbb{R}}& \exp\bigg\{ -\frac{n}{2}s^2 \bigg\} \\
&\times \prod\limits_{j = 1}^n \exp\left\{ -s\sqrt{4na_2}\cdot \overline{\psi}_{j1}\psi_{j1}\overline{\psi}_{j2}\psi_{j2} \right\} ds,
\end{align*}
where $\mathcal{H}_2$ is the space of self-adjoint operators in $\operatorname{End} \mathbb{C}^{2}$ and
\begin{align*}
Q &= \left( \begin{array}{cc}
t_1 & u + iv \\
u - iv & t_{2}
\end{array} \right); \\
dQ &= dt_1 dt_2 dudv.
\end{align*}
Set $b_2 = -\sqrt{4na_2} = -\sqrt{\frac{2(n-p)}{pn}}$. Now we can integrate over the Grassmann variables
\begin{equation*}
F_2(\Lambda) = 2\bigg(\frac{n}{2\pi}\bigg)^{5/2}\int\limits_{\mathbb{R}} e^{-\frac{n}{2}s^2} \int\limits_{\mathcal{H}_2} (b_2 s - \det{(Q - i\Lambda)})^n e^{-\frac{n}{2}\operatorname{tr}{Q^2}}dQ ds
\end{equation*}

Change the variables $t_j \rightarrow t_j + i\lambda_j$ and move the line of integration back to the real axis. Indeed, consider the rectangular contour with vertices in the points $(-R, 0)$, $(R, 0)$, $(R, -i\lambda_j)$ and $(-R, -i\lambda_j)$. Since the integrand is a holomorphic on $\mathbb{C}$ function, the integral over this contour is zero. Because the integrand is a polynomial multiplied by exponent, the integral over the vertical sides of the contour tends to 0 when $R \to \infty$. So, recalling that $\lambda_j = \lambda_0 + x_j/n$, we can write 
\begin{multline*}
F_2(\Lambda) = \frac{C_n(X)}{\pi} \cdot e^{\lambda_0(x_1+x_2)} \\
\times \int\limits_{\mathbb{R}} e^{-\frac{n}{2}s^2} \int\limits_{\mathcal{H}_2} (b_2 s - \det{Q})^n e^{-\frac{n}{2}\operatorname{tr}{(Q+i\Lambda_0)^2}} \exp\{-i\tr XQ \}dQ ds,
\end{multline*}
where 
\begin{equation}
\label{C_n_definition}
C_n(X) = n\bigg(\frac{n}{2\pi}\bigg)^{3/2} e^{\frac{1}{2n}(x_1^2+x_2^2)}.
\end{equation}

Let us change the variables $Q \to U^*TU$, where $U$ is a unitary matrix and $T = \operatorname{diag}\{t_1, t_2\}$. Then $dQ$ changes to $\frac{\pi}{2}(t_1 - t_2)^2 dt_1 dt_2$ and
\begin{multline*}
F_2(\Lambda) = \frac{1}{2} C_n(X) e^{\lambda_0(x_1+x_2)}\int\limits_{\mathbb{R}} e^{-\frac{n}{2}s^2} \int\limits (t_1 - t_2)^2 (b_2 s - t_1t_2)^n \\
\exp\bigg\{-\frac{n}{2}\sum\limits_{j=1}^2 (t_j+i\lambda_0)^2\bigg\} \int\limits_{U_2} e^{-i\operatorname{tr}{(XU^{*}TU)}}dU_2(U) dT ds,
\end{multline*}
where $U_2$ is the group of the unitary $2 \times 2$ matrices, $dU_2(U)$ is the normalized to unity Haar measure, $dT = dt_1 dt_2$.

The integration over the unitary group using the Harish-Chandra/Itsykson--Zuber formula \eqref{eq:H-C/I--Z (symm.domain)} implies the assertion of Remark \ref{rem1}.


\subsubsection{General case $m > 1$}

Let us transform the exponent of \eqref{after_averaging_m>1}.
\begin{align*}
(\chi_{jk}\chi_{kj})^l &= (l!)^2 \sum\limits_{\alpha, \beta \in I_{2m, l}} \prod\limits_{q = 1}^l \overline{\psi}_{j\alpha_q}\psi_{k\alpha_q}\overline{\psi}_{k\beta_q}\psi_{j\beta_q} \\
&= (-1)^l (l!)^2 \sum\limits_{\alpha, \beta \in I_{2m, l}} \prod\limits_{q = 1}^l \overline{\psi}_{j\alpha_q}\psi_{j\beta_q} \prod\limits_{q = 1}^l \overline{\psi}_{k\beta_q}\psi_{k\alpha_q},
\end{align*}
where $I_{2m, l}$ is defined in \eqref{eq:I_nk definition}.

Since $\psi_{jl}^2 = \overline{\psi}_{jl}^2 = 0$, we have
\begin{align*}
\sum\limits_{j < k} (\chi_{jk}\chi_{kj})^l + \sum\limits_{j=1}^n \frac{1}{2}\chi_{jj}^{2l} &= (-1)^l (l!)^2 \bigg(\sum\limits_{j < k} \sum\limits_{\alpha, \beta \in I_{2m,l}} \prod\limits_{q = 1}^l \overline{\psi}_{j\alpha_q}\psi_{j\beta_q} \prod\limits_{q = 1}^l \overline{\psi}_{k\beta_q}\psi_{k\alpha_q} \\
&+ \frac{1}{2}\sum\limits_{j = 1}^n \prod\limits_{q = 1}^l \overline{\psi}_{j\alpha_q}\psi_{j\beta_q} \prod\limits_{q = 1}^l \overline{\psi}_{j\beta_q}\psi_{j\alpha_q}\bigg)\\
&= \frac{1}{2} (-1)^l (l!)^2 \sum\limits_{\alpha, \beta \in I_{2m, l}}\sum\limits_{j, k} 
\prod\limits_{q = 1}^l \overline{\psi}_{j\alpha_q}\psi_{j\beta_q} \prod\limits_{q = 1}^l \overline{\psi}_{k\beta_q}\psi_{k\alpha_q} \\
&= \frac{1}{2} (-1)^l (l!)^2 \sum\limits_{\alpha, \beta}\bigg(\sum\limits_{j} 
\prod\limits_{q = 1}^l \overline{\psi}_{j\alpha_q}\psi_{j\beta_q}\bigg)\bigg(\sum\limits_{k} \prod\limits_{q = 1}^l \overline{\psi}_{k\beta_q}\psi_{k\alpha_q}\bigg) \\
&= (-1)^l (l!)^2 \bigg(\frac{1}{2}\sum\limits_\alpha \bigg( \sum\limits_{j = 1}^n \prod\limits_{q = 1}^l \overline{\psi}_{j\alpha_q}\psi_{j\alpha_q} \bigg)^2 \\
&+ \sum\limits_{\alpha \prec \beta} \bigg(\sum\limits_{j = 1}^n \prod\limits_{q = 1}^l \overline{\psi}_{j\alpha_q}\psi_{j\beta_q} \bigg)\bigg(\sum\limits_{j = 1}^n \prod\limits_{q = 1}^l \overline{\psi}_{j\beta_q}\psi_{j\alpha_q} \bigg)\bigg),
\end{align*}
where $\prec$ is the lexicographical order.

Hubbard-Stratonovich transformation \eqref{sqrt_of_exponent} 
yields
\begin{align*}
\exp\bigg\{(-1)^l (l!)^2 \frac{a_l}{2} \bigg( \sum\limits_{j = 1}^n \prod\limits_{q = 1}^l \overline{\psi}_{j\alpha_q}\psi_{j\alpha_q} &\bigg)^2\bigg\}  \\
= \sqrt{\frac{n}{2\pi}}\int \exp\bigg\{b_l (Q_l)_{\alpha\alpha}& \bigg( \sum\limits_{j = 1}^n \prod\limits_{q = 1}^l \overline{\psi}_{j\alpha_q}\psi_{j\alpha_q} \bigg) - \frac{n}{2}(Q_l)_{\alpha\alpha}^2\bigg\} d(Q_l)_{\alpha\alpha}, \\
\exp\bigg\{(-1)^l (l!)^2 a_l \bigg(\sum\limits_{j = 1}^n \prod\limits_{q = 1}^l \overline{\psi}_{j\alpha_q}\psi_{j\beta_q} &\bigg)\bigg(\sum\limits_{j = 1}^n \prod\limits_{q = 1}^l \overline{\psi}_{j\beta_q}\psi_{j\alpha_q} \bigg)\bigg\} \\
= \frac{n}{\pi}\int \exp\{-n|(Q_l)_{\alpha\beta}|^2\}& \exp\bigg\{b_l\bigg(\sum\limits_{j = 1}^n \prod\limits_{q = 1}^l \overline{\psi}_{j\alpha_q}\psi_{j\beta_q} \bigg)(Q_l)_{\alpha\beta} \\
{} + b_l\bigg(\sum\limits_{j = 1}^n& \prod\limits_{q = 1}^l \overline{\psi}_{j\beta_q}\psi_{j\alpha_q} \bigg)\overline{(Q_l)_{\alpha\beta}}\bigg\} d\Re (Q_l)_{\alpha\beta} d\Im (Q_l)_{\alpha\beta},
\end{align*}
where
\begin{equation}\label{eq:b_l definition}
b_l = i^l l!\sqrt{na_l}.
\end{equation}
The above computations compose into the following representation of the exponent of \eqref{after_averaging_m>1}
\begin{multline}\label{eq:transformation of exp}
\exp\bigg\{ a_l\bigg(\sum\limits_{j < k} (\chi_{jk}\chi_{kj})^l + \frac{1}{2} \sum\limits_{j=1}^n \chi_{jj}^{2l}\bigg) \bigg\} = \left( \frac{1}{2} \right)^{\frac{1}{2}\binom{2m}{l}} \bigg( \frac{n}{\pi} \bigg)^{\frac{1}{2}\binom{2m}{l}^2} \\
\times \int\limits_{\mathcal{H}_{2m, l}} \exp\left\{ -\frac{n}{2}\operatorname{tr}Q_l^2 \right\}\prod\limits_{j = 1}^n \exp\bigg\{ b_l\sum\limits_{\alpha, \beta} (Q_l)_{\alpha\beta} \prod\limits_{q = 1}^l \overline{\psi}_{j\alpha_q}\psi_{j\beta_q} \bigg\} dQ_l 
\end{multline}
where $\mathcal{H}_{2m, l}$ is the space of self-adjoint operators in $\operatorname{End} \Lambda^l \mathbb{C}^{2m}$ and $dQ_l$ is defined in \eqref{eq:measure on H_2m,l}.
Therefore, substitution of \eqref{eq:transformation of exp} into \eqref{after_averaging_m>1} gives us
\begin{multline}\label{eq:integral over H_m and Grassmanns}
F_{2m}(\Lambda) = Z_n^{(2m)}\int\limits_{\mathcal{H}_{2m, 1} \times H_m} \prod\limits_{l = 1}^{2m} \exp\left\{ -\frac{n}{2}\operatorname{tr} Q_l^2 \right\} \prod\limits_{j = 1}^n \exp\bigg\{ b_l \sum\limits_{\alpha, \beta} (Q_l)_{\alpha\beta} \prod\limits_{q = 1}^l \overline{\psi}_{j\alpha_q}\psi_{j\beta_q} \\
- \left(\frac{1}{2} - \frac{1}{2^l}\right)a_l \chi_{jj}^{2l} + \lambda_l\overline{\psi}_{jl}\psi_{jl} \bigg\}d\overline{\psi}_{jl}d\psi_{jl} \prod\limits_{l = 1}^{2m} dQ_l,
\end{multline}
where $H_m$ is defined in \eqref{eq:H_m} and
\begin{equation}
\notag 
Z_n^{(2m)} = \left( \frac{1}{2} \right)^{\frac{1}{2}(2^{2m} - 1)} \bigg( \frac{n}{\pi} \bigg)^{\frac{1}{2}\left(\binom{4m}{2m} - 1\right)}.
\end{equation}

Now we can expand the exponents of \eqref{eq:integral over H_m and Grassmanns} into the series
\begin{multline}\label{eq:exp expansion}
\exp\bigg\{ \sum\limits_{l = 1}^{2m} b_l\sum\limits_{\alpha, \beta} (Q_l)_{\alpha\beta} \prod\limits_{q = 1}^l \overline{\psi}_{j\alpha_q}\psi_{j\beta_q} - \left(\frac{1}{2} - \frac{1}{2^l}\right)a_l \chi_{jj}^{2l} + \lambda_l\overline{\psi}_{jl}\psi_{jl} \bigg\} \\
= \exp\bigg\{ \sum\limits_{l = 1}^{2m}\sum\limits_{\alpha, \beta}  (\widetilde{Q}_l)_{\alpha\beta} \prod\limits_{q = 1}^l \overline{\psi}_{j\alpha_q}\psi_{j\beta_q} \bigg\} = \sum\limits_{k = 1}^{2m} \frac{1}{k!}\bigg(\sum\limits_{l = 1}^{2m}\sum\limits_{\alpha, \beta}  (\widetilde{Q}_l)_{\alpha\beta} \prod\limits_{q = 1}^l \overline{\psi}_{j\alpha_q}\psi_{j\beta_q}\bigg)^k,
\end{multline}
where 
\begin{equation*}
\begin{array}{cc}
\widetilde{Q}_1 = b_1Q_1 + \Lambda, &\quad \widetilde{Q}_l = b_lQ_l - \tilde{b}_lI, \\
\tilde{b}_{2l} = (2^{-1} - 2^{-l})a_{l}, &\quad \tilde{b}_{2l - 1} = 0.
\end{array}
\end{equation*}
The most important terms contain all $4m$ Grassmann variables $\{\psi_{js},\, \overline{\psi}_{js}\}_{s = 1}^{2m}$, because the other terms become zeros after integration over Grassmann variables. Thus, expansion of \eqref{eq:exp expansion} with Lemma \ref{lemma_1} implies
\begin{multline*} 
\int\exp\bigg\{ \sum\limits_{l = 1}^{2m}\sum\limits_{\alpha, \beta}  (\widetilde{Q}_l)_{\alpha\beta} \prod\limits_{q = 1}^l \overline{\psi}_{j\alpha_q}\psi_{j\beta_q} \bigg\} \prod\limits_{l = 1}^{2m} d\overline{\psi}_{jl}d\psi_{jl} \\
= \int\sum\limits_{\substack{k_1 + 2k_2 + \ldots + 2mk_{2m} = 2m \\ k_j \in \mathbb{Z}_+}} \frac{1}{(k_1 + \ldots + k_{2m})!} \cdot \frac{(k_1 + \ldots + k_{2m})!}{k_1! \ldots k_{2m}!} \cdot \frac{(2m)!}{(1!)^{k_1} \ldots ((2m)!)^{k_{2m}}} \\
\times \sum\limits_{\alpha, \beta \in I_{2m, 2m}} \bigg(\bigwedge\limits_{s = 1}^{2m} \widetilde{Q}_s^{\wedge k_s}\bigg)_{\alpha\beta} \prod\limits_{q = 1}^{2m} \overline{\psi}_{j\alpha_q}\psi_{j\beta_q} \prod\limits_{l = 1}^{2m} d\overline{\psi}_{jl}d\psi_{jl},
\end{multline*}
where only the most important terms remain. Integration over the Grassmann variables and substitution of the result into \eqref{eq:integral over H_m and Grassmanns} gives us
\begin{equation*}
F_{2m}(\Lambda) = Z_n^{(2m)}\int\limits_{H_m}\int\limits_{\mathcal{H}_{2m, 1}} \left(A_{2m}(\hat{Q}_1, Q)\right)^n \exp\bigg\{-\frac{n}{2}\sum\limits_{l = 1}^{2m} \operatorname{tr}{Q_l^2}\bigg\}dQ_1 dQ,
\end{equation*}
where $\hat{Q}_1 = Q_1 + \frac{1}{b_1}\Lambda = Q_1 - i\Lambda$, $Q = (Q_2, \ldots, Q_{2m})$, $dQ = \prod\limits_{l = 2}^{2m} dQ_l$ and $A_{2m}$ is defined in \eqref{eq:A_2m definition}.

Change the variables $(Q_1)_{jj} \rightarrow (Q_1)_{jj} + i\lambda_j$ and move the line of integration back to the real axis. Similarly to the case $m = 1$, the Cauchy theorem yields
\begin{multline*}
F_{2m}(\Lambda) = \frac{C_n^{(2m)}(X)}{\pi^m} \cdot \exp\bigg\{\lambda_0\sum\limits_{j = 1}^{2m} x_j\bigg\} \int\limits_{H_{m}}\int\limits_{\mathcal{H}_{2m, 1}} \left(A_{2m}(Q_1, Q)\right)^n \\
\times \exp\bigg\{-\frac{n}{2}\bigg( \tr(Q_1 + i\Lambda_0)^2 + \sum\limits_{l = 2}^{2m} \tr{Q_l^2} \bigg)\bigg\} e^{-i\tr XQ_1}dQ_1 dQ,
\end{multline*}
where $C_n^{(2m)}(X)$ is defined in \eqref{C_n^(2m)_definition}.
Let us change the variables $Q_1 = U^*TU$, where $U$ is a unitary operator and $T = \operatorname{diag}\{t_j\}_{j = 1}^{2m}$. Then $dQ$ changes to $\pi^m K_{2m}^{-1}\Delta(T)^2 dT$ and
\begin{multline}
\label{F_2m_variables_change_0}
F_{2m}(\Lambda) = \frac{C_n^{(2m)}(X)}{K_{2m}} \cdot \exp\bigg\{\lambda_0\sum\limits_{j = 1}^{2m} x_j\bigg\} \int\limits_{H_m} \int\limits_{\mathbb{R}^{2m}} \Delta(T)^2 \left(A_{2m}(Q_1, Q)\right)^n \\
\times \exp\bigg\{-\frac{n}{2}\bigg(\sum\limits_{j=1}^{2m} (t_j + i\lambda_0)^2 + \sum\limits_{l = 2}^{2m} \operatorname{tr}{Q_l^2} \bigg)\bigg\} \int\limits_{U_{2m}} e^{-i\operatorname{tr} XU^{*}TU}dU_{2m}(U) dT dQ,
\end{multline}
where $K_{2m} = \prod\limits_{j = 1}^{2m} j!$, $\Delta(T)$ is defined in \eqref{eq:Vandermonde definition}, $U_{2m}$ is the subgroup of the unitary operators in $\End \mathbb{C}^{2m}$, $dU_{2m}(U)$ is the normalized to unity Haar measure, $dT = \prod\limits_{j = 1}^{2m} dt_j$. Transform $A_{2m}(Q_1, Q)$
\begin{multline*}
(b_1Q_{1})^{\wedge k_1}\wedge \bigwedge\limits_{s = 2}^{2m} (b_sQ_{s} - \tilde{b}_sI)^{\wedge k_s} 
= (U^*b_1TU)^{\wedge k_1}\wedge \bigwedge\limits_{s = 2}^{2m} ((U^*U)^{\wedge s}(b_sQ_{s} - \tilde{b}_sI)(U^*U)^{\wedge s})^{\wedge k_s},
\end{multline*}
where $I$ is the identity operator.
The assertion (iv) of Proposition \ref{proposition_1} implies
\begin{multline*}
(U^*b_1TU)^{\wedge k_1}\wedge \bigwedge\limits_{s = 2}^{2m} ((U^*U)^{\wedge s}(b_sQ_{s} - \tilde{b}_sI)(U^*U)^{\wedge s})^{\wedge k_s} \\
= (U^*)^{\wedge 2m} \bigg((b_1T)^{\wedge k_1}\wedge \bigwedge\limits_{s = 2}^{2m} (b_sU^{\wedge s}Q_{s}(U^*)^{\wedge s} - \tilde{b}_sI)^{\wedge k_s}\bigg) U^{\wedge 2m} \\
= (b_1T)^{\wedge k_1}\wedge \bigwedge\limits_{s = 2}^{2m} (b_sU^{\wedge s}Q_{s}(U^*)^{\wedge s} - \tilde{b}_sI)^{\wedge k_s}.
\end{multline*}

Change the variables $Q_l = (U^*)^{\wedge l}R_l U^{\wedge l}$, $l = \overline{2, 2m}$. Since $U^{\wedge l}$ is the unitary operator
, $dQ$ changes to $dR = \prod\limits_{l = 2}^{2m} dR_l$. Then \eqref{F_2m_variables_change_0} implies
\begin{multline*}
F_{2m}(\Lambda) = \frac{C_n^{(2m)}(X)}{K_{2m}} \cdot \exp\bigg\{\lambda_0\sum\limits_{j = 1}^{2m} x_j\bigg\} \int\limits_{H_m} \int\limits_{\mathbb{R}^{2m}} \Delta(T)^2 \left(A_{2m}(T, R)\right)^n \\
\times \exp\bigg\{-\frac{n}{2}\bigg(\sum\limits_{j=1}^{2m} (t_j + i\lambda_0)^2 + \sum\limits_{l = 2}^{2m} \operatorname{tr}{R_l^2} \bigg)\bigg\} \int\limits_{U_{2m}} e^{-i\operatorname{tr} XU^{*}TU}dU_{2m}(U)dT dR,
\end{multline*}
where $R = (R_2, \ldots, R_{2m})$.
\begin{equation}\label{eq:symm func}
\int\limits_{H_m} \left(A_{2m}(T, R)\right)^n \exp\bigg\{-\frac{n}{2}\bigg(\sum\limits_{j=1}^{2m} (t_j + i\lambda_0)^2 + \sum\limits_{l = 2}^{2m} \operatorname{tr}{R_l^2} \bigg)\bigg\} dR
\end{equation}
is a symmetric function of $\{t_j\}_{j = 1}^{2m}$. Indeed, after swapping $t_{j_1}$ and $t_{j_2}$ and changing the variables $R_l \rightarrow (\mathcal{M}_{j_1j_2}^*)^{\wedge l}R_l\mathcal{M}_{j_1j_2}^{\wedge l}$, where $\mathcal{M}_{j_1j_2}$ is the unit matrix, in which rows $j_1$ and $j_2$ are swapped, the integrand in \eqref{eq:symm func} remains unchanged. Hence, the integration over the unitary group using the Harish-Chandra/Itsykson--Zuber formula \eqref{eq:H-C/I--Z (symm.domain)} can be done, which yields the assertion of Proposition \ref{pr1(integral representation)}.
%

\section{Proof of Theorem \ref{th_1}}\label{sec:th1}
To find asymptotics of $F_2(\Lambda)$, we apply the steepest descent method to the integral representation \eqref{final_integral_representation}. As usual, the key technical point of the steepest descent method is to choose a good contour of integration (in our case it is 3 dimension space of $(t_1, t_2, s)$), which contains  the stationary point $(t_1^*, t_2^*, s^*)$ of $f$ and then to prove that for any $(t_1, t_2, s)$ in our ``contour''
\begin{equation}\label{eq:sdm inequality}
\Re f(t_1, t_2, s) \leq \Re f(t_1^*, t_2^*, s^*).
\end{equation}
Let us introduce the function $h_\alpha: \mathbb{R}^5 \to \mathbb{R}$
\begin{equation}
\label{h_alpha_definition}
h_{\alpha}(t_1, t_2, s, b_2, \lambda_0) = \frac{1}{2}\bigg(\log A - \sum\limits_{j=1}^2 t_j^2 - s^2 - \left(\frac{1 - \alpha}{\alpha}\right)^2 b_2^2 - 2\alpha(1 - \alpha)\lambda_0^2 \bigg),
\end{equation}
where 
\begin{equation}
\label{A_and_B}
A = \left(b_2 s - t_1 t_2 + \alpha^2\lambda_0^2\right)^2 + \alpha^2\lambda_0^2(t_1 + t_2)^2
.
\end{equation} 
Then $\Re f$ at our ``contour'' (which is defined further) has the form
\begin{equation*}
\Re f\left(T, s\right) = h_{\alpha}(\Re t_1, \Re t_2, s, b_2, \lambda_0) + \frac{1}{2}\left(\frac{1-\alpha}{\alpha}\right)^2 b_2^2 + (1-\alpha)\lambda_0^2
\end{equation*}
for some $\alpha$. To prove \eqref{eq:sdm inequality}, we use the following lemma.

\begin{lemma}
\label{lemma_2}
Let $h_\alpha$ be defined by \eqref{h_alpha_definition} and \eqref{A_and_B}. Then for every $\alpha \in [1/2,\,1)$, $t_1$, $t_2$, $s$, $b_2$, $\lambda_0 \in \mathbb{R}$ the following inequality holds
\begin{equation}
\label{lemma's_inequality}
h_{\alpha}(t_1, t_2, s, b_2, \lambda_0) \leq \frac{1}{2}\log \left(\frac{\alpha}{1 - \alpha}\right)^2 - 1
\end{equation}
Moreover, the equality holds if and only if at least one of the following conditions is satisfied
\begin{enumerate}[(a)]
\item $\alpha = 1/2,\, t_1 = -t_2 = \pm \sqrt{4 - 4b_2^2 - \lambda_0^2}/2,\, s = b_2$;
\item $\alpha = 1/2,\, t_1 = t_2 = \pm \sqrt{4 - 4b_2^2 - \lambda_0^2}/2,\, s = -b_2,\, b_2\lambda_0 = 0$;
\item $t_1 = t_2 = 0,\, s = b_2\frac{1-\alpha}{\alpha},\, \alpha(1-\alpha)\lambda_0^2+\left(\frac{1-\alpha}{\alpha}\right)^2b_2^2 = 1$;
\item $t_1 = t_2 = 0,\, s = -b_2\frac{1-\alpha}{\alpha},\, b_2 = \pm \frac{\alpha}{1 - \alpha},\, \lambda_0 = 0$.
\end{enumerate}
\end{lemma}

\emph{Proof.} Rewrite the inequality \eqref{lemma's_inequality} in the form
\begin{equation*}
\log\frac{1 - \alpha}{\alpha}A^{1/2} + 1 \le \frac{1}{2}\left(t_1^2 + t_2^2 + s^2 + d^2 + 2\alpha(1 - \alpha)\lambda_0^2\right),
\end{equation*}
where $d = \frac{1 - \alpha}{\alpha}b_2$. Since 
\begin{equation}\label{eq:log's convexity}
\log\frac{1 - \alpha}{\alpha}A^{1/2} \le \frac{1 - \alpha}{\alpha}A^{1/2} - 1,
\end{equation}
it is sufficient to prove
\begin{equation} \label{eq:lemma's_inequality 2}
\left(\frac{1 - \alpha}{\alpha}\right)^2 A \le \frac{1}{4}\left(t_1^2 + t_2^2 + s^2 + d^2 + 2\alpha(1 - \alpha)\lambda_0^2\right)^2.
\end{equation}
Recalling \eqref{A_and_B}, we have
\begin{multline*}
\left(\frac{1 - \alpha}{\alpha}\right)^2 A = s^2d^2 + \left(\frac{1 - \alpha}{\alpha}\right)^2 t_1^2t_2^2 + \alpha^2(1 - \alpha)^2\lambda_0^4 \\
- 2\frac{1 - \alpha}{\alpha}sdt_1t_2 + 2\alpha(1 - \alpha)\lambda_0^2sd + (1 - \alpha)^2\lambda_0^2(t_1^2 - t_2^2).
\end{multline*}
\eqref{eq:lemma's_inequality 2} is transformed into
\begin{multline*}
s^2d^2 + \left(\frac{1 - \alpha}{\alpha}\right)^2 t_1^2t_2^2 - 2\frac{1 - \alpha}{\alpha}sdt_1t_2 + 2\alpha(1 - \alpha)\lambda_0^2sd \\
+ (1 - \alpha)^2\lambda_0^2(t_1^2 - t_2^2) \le \frac{1}{4}(t_1^2 + t_2^2)^2 + \frac{1}{4}(s^2 + d^2)^2 \\
+ \frac{1}{2}(t_1^2 + t_2^2)(s^2 + d^2) + \alpha(1 - \alpha)\lambda_0^2(t_1^2 + t_2^2) + \alpha(1 - \alpha)\lambda_0^2(s^2 + d^2).
\end{multline*}
The last inequality is the sum of following obvious inequalities
\begin{align}
\label{eq:alpha}
(1 - \alpha)^2\lambda_0^2(t_1^2 - t_2^2) &\le \alpha(1 - \alpha)\lambda_0^2(t_1^2 + t_2^2), \\
\label{eq:s^2d^2}
s^2d^2 &\le \frac{1}{4}(s^2 + d^2)^2, \\
\label{eq:t_1^2t_2^2}
\left(\frac{1 - \alpha}{\alpha}\right)^2 t_1^2t_2^2 &\le t_1^2t_2^2 \le \frac{1}{4}(t_1^2 + t_2^2)^2, \\
\label{eq:sdt_1t_2}
- 2\frac{1 - \alpha}{\alpha}sdt_1t_2 &\le 2|sdt_1t_2| \le \frac{1}{2}(t_1^2 + t_2^2)(s^2 + d^2), \\
\label{eq:sd}
2\alpha(1 - \alpha)\lambda_0^2sd &\le \alpha(1 - \alpha)\lambda_0^2(s^2 + d^2).
\end{align}

It remains to determine conditions when the equality in \eqref{lemma's_inequality} holds. It holds if and only if the equalities in \eqref{eq:log's convexity}, \eqref{eq:alpha}-\eqref{eq:sd} hold. Let $(n')$ denotes the corresponding equality for inequality $(n)$. Then
\begin{align}
\label{eq:main equation}
(\ref{eq:log's convexity}') &\Leftrightarrow A^{1/2} = \frac{\alpha}{1 - \alpha}, \\
\notag
(\ref{eq:s^2d^2}') &\Leftrightarrow s^2 = d^2, \\
\notag
(\ref{eq:t_1^2t_2^2}') &\Rightarrow t_1^2 = t_2^2.
\end{align}
Everywhere below until the end of the proof we assume that $s^2 = d^2$ and $t_1^2 = t_2^2$. Then
\begin{align*}
(\ref{eq:alpha}') &\Leftrightarrow \left[
\begin{array}{l}
\alpha = 1/2; \\
\lambda_0t_1 = 0;
\end{array}
\right. \\
(\ref{eq:t_1^2t_2^2}') &\Leftrightarrow \left[
\begin{array}{l}
\alpha = 1/2; \\
t_1 = 0;
\end{array}
\right. \\
(\ref{eq:sdt_1t_2}') &\Leftrightarrow \left[
\begin{array}{l}
\begin{cases}
\alpha = 1/2; \\
sdt_1t_2 \le 0;
\end{cases} \\
t_1s = 0;
\end{array}
\right. \\
(\ref{eq:sd}') &\Leftrightarrow \left[
\begin{array}{l}
sd \ge 0; \\
\lambda_0 = 0.
\end{array}
\right.
\end{align*}

Let us consider the following cases
\begin{enumerate}
	\item $t_1 = 0$.
	\begin{enumerate}
		\item $\lambda_0 = 0$. Then \eqref{eq:main equation} is transformed into $(b_2s)^2 = \big(\frac{\alpha}{1 - \alpha}\big)^2$. Since $s^2 = d^2$, we get $b_2 = \pm \frac{\alpha}{1 - \alpha}$ that implies (d).
		\item $sd \ge 0$. Then \eqref{eq:main equation} is equivalent to $\alpha^2\lambda_0^2+\frac{1-\alpha}{\alpha}b_2^2 = \frac{\alpha}{1 - \alpha}$ that implies (c).
	\end{enumerate}
	\item $t_1 \ne 0 \Rightarrow \alpha = 1/2$.
	\begin{enumerate}
		\item $s = 0$. Hence, \eqref{eq:main equation} is transformed into $\lambda_0^2/4 + t_1^2 = 1$ that implies (b).
		\item $s \ne 0 \Rightarrow dst_1t_2 < 0$.
		\begin{enumerate}
			\item $sd > 0$. Then \eqref{eq:main equation} is transformed into $\lambda_0^2/4 + b_2^2 + t_1^2 = 1$. Condition (a) is satisfied.
			\item $sd < 0 \Rightarrow \lambda_0 = 0$. Then \eqref{eq:main equation} is transformed into $b_2^2 + t_1^2 = 1$. Condition (b) is satisfied.
		\end{enumerate}
	\end{enumerate}
\end{enumerate}

Finally, it is easy to check that the values of $h_{\alpha}$ at the points satisfying (a)-(d) are equal to the r.h.s. of \eqref{lemma's_inequality}. $\blacksquare$

\medskip
Now we are ready to prove Theorem \ref{th_1}. We start from the lemma 
\begin{lemma}\label{lem:1.1}
Let all conditions of Theorem \ref{th_1} are hold and $\lambda_0 \in (\lambda_*(p), \lambda_*(p))$. Then $F_2(\Lambda)$ satisfies the asymptotic relation
\begin{multline}
\label{asymptotics_for_F_case_1}
F_2(\Lambda) = 2n \exp\{n(\lambda_0^2+b_2^2-2)/2 + \lambda_0(x_1 + x_2)/2\} \\
\times \frac{\sin((x_1-x_2)\sqrt{4-4b_2^2-\lambda_0^2}/2)}{(x_1-x_2)}(1 + o(1)),
\end{multline}
where $b_2$ is defined in \eqref{eq:b_l definition}.
\end{lemma}
\emph{Proof.} Set
\begin{align}
\notag
t_\eta^{*} &= \frac{(-1)^\eta}{2} \sqrt{4 - 4b_2^2 - \lambda_0^2}; \\
\label {T^(jk)_definition}
T_*^{(\eta\nu)} &= \diag\left\{ t_\eta^*, t_\nu^* \right\} - i\Lambda_0/2,
\end{align}
where $\eta, \nu = 1, 2$.

Consider the contour $\Im t_1 = \Im t_2 = -\lambda_0/2,\, s \in \mathbb{R}$. It contains the points $(t_1^* - i\lambda_0/2, t_2^* - i\lambda_0/2, b_2)$ and $(t_2^* - i\lambda_0/2, t_1^* - i\lambda_0/2, b_2)$, which are the stationary points of $f$. The contour may contain another stationary points of $f$, but this fact does not affect the proof, except the case $\lambda_0 = 0$ for which the points $(t_1^*, t_1^*, -b_2)$ and $(t_2^*, t_2^*, -b_2)$ are also under consideration. First, consider the case $\lambda_0 \ne 0$.

Shift the variables $t_\eta \rightarrow t_\eta - i\lambda_0/2$ in \eqref{final_integral_representation} and restrict the integration domain by
\begin{equation*}
B_R = \left\{(t_1, t_2, s) \in \mathbb{R}^3: \max\{|t_1|, |t_2|, |s|\} \leq R\right\}.
\end{equation*}
Then
\begin{equation}
\label{integral_without_tails}
F_2(\Lambda) = C_n(X) \frac{ie^{\lambda_0(x_1 + x_2)/2}}{(x_1 - x_2)} \int\limits_{B_R} g(T, s)e^{n f\left(T - \frac{i}{2}\Lambda_0, s\right)} dT ds + O(e^{-nR^2/4}),\, n \to \infty,
\end{equation}
where $f$ is defined in \eqref{f_definition} and
\begin{align*}
g(T, s) &= (t_1 - t_2) \exp\bigg\{-i\sum\limits_{\eta=1}^2 x_\eta t_\eta\bigg\}.
\end{align*}

Then it is easy to see, that for $\eta \ne \nu$
\begin{align*}
f(T_*^{(\eta\nu)}, b_2) &= \frac{1}{2}b_2^2 + \frac{1}{2}\lambda_0^2 - 1; \\
f''(T_*^{(\eta\nu)}, b_2) &= -\left( \begin{array}{ccc}
(t_\nu^{*} - i\lambda_0/2)^2 + 1 & b_2^2 & -b_2 (t_\nu^{*} - i\lambda_0/2) \\
b_2^2 & (t_\eta^{*} - i\lambda_0/2)^2 + 1 & -b_2 (t_\eta^{*} - i\lambda_0/2) \\
-b_2 (t_\nu^{*} - i\lambda_0/2) & -b_2 (t_\eta^{*} - i\lambda_0/2) & b_2^2 + 1
\end{array} \right); \\
\det f''(T_*^{(\eta\nu)}, b_2) &= -(4 - 4b_2^2 - \lambda_0^2) < -(\lambda_*(p)^2 - \lambda_0^2); \\
\Re f''(T_*^{(\eta\nu)}, b_2) &= -\left( \begin{array}{ccc}
(t_\nu^{*})^2 - \lambda_0^2/4 + 1 & b_2^2 & -b_2 t_\nu^{*} \\
b_2^2 & (t_\eta^{*})^2 - \lambda_0^2/4 + 1 & -b_2 t_\eta^{*} \\
-b_2 t_\nu^{*} & -b_2 t_\eta^{*} & b_2^2 + 1
\end{array} \right); \\
\det \Re f''(T_*^{(\eta\nu)}, b_2) &= -\left(1 - \lambda_0^2/4\right)(4 - 4b_2^2 - \lambda_0^2) < -(\lambda_*(p)^2 - \lambda_0^2)^2/4,
\end{align*}
where $T_*^{(\eta\nu)}$ is defined in \eqref{T^(jk)_definition}.

Note that 
\begin{equation*}
\Re f\left(T - \frac{i}{2}\Lambda_0, s\right) = h_{1/2}(t_1, t_2, s, b_2, \lambda_0) + \frac{1}{2}b_2^2 + \frac{1}{2}\lambda_0^2,
\end{equation*}
where $h_\alpha$ is defined in \eqref{h_alpha_definition}. According to Lemma \ref{lemma_2}, $\Re f\left(T - \frac{i}{2}\Lambda_0, s\right)$, as a function of real variables $t_1$, $t_2$, $s$, attains its maximum at $(T_*^{(\eta\nu)}, b_2)$. Hence, $\Re f''(T_*^{(\eta\nu)}, b_2)$ is nonpositive, but since $\det \Re f''(T_*^{(\eta\nu)}, b_2) < 0$, $\Re f''(T_*^{(\eta\nu)}, b_2)$ is negative definite.

Let $V_n^{(\eta\nu)}$ be a $n^{-1/2} \log n\text{-neighborhood}$ of the point $(T_*^{(\eta\nu)}, b_2)$ and let $V_n$ everywhere below denote the union of such neighborhoods of the stationary points under consideration, unless otherwise stated. Then for $\left(T, s\right) \notin V_n$ and sufficiently large $n$ we have
\begin{multline*}
\Re f(T_*^{(\eta\nu)}, b_2) - \Re f\left(T - \frac{i}{2}\Lambda_0, s\right) \\
\geq \min_{\eta \ne \nu} \min_{\left(T - \frac{i}{2}\Lambda_0, s\right) \in \partial V_n^{(\eta\nu)}} \left\{ \Re f(T_*^{(\eta\nu)}, b_2) - \Re f\left(T - \frac{i}{2}\Lambda_0, s\right) \right\} \geq C\frac{\log^2 n}{n},
\end{multline*}

Thus we can restrict the integration domain by $V_n$.

Set $q = (t_1, t_2, s),\, q^{*} = (t_\eta^{*}, t_\nu^{*}, b_2).$ Then expanding $f$ and $g$ by the Taylor formula and changing the variables $q \rightarrow n^{-1/2}q + q^{*}$, we get
\begin{multline*}
F_2(\Lambda) = n^{-3/2} C_n(X) \cdot \frac{i \exp\{n(\lambda_0^2+b_2^2-2)/2 + \lambda_0(x_1 + x_2)/2\}}{(x_1 - x_2)} \\
\times \bigg(\sum\limits_{\eta \ne \nu} \int\limits_{[-\log n, \log n]^3} g(\Re T_*^{(\eta\nu)}, b_2) \exp\left\{\frac{1}{2} q f''(T_*^{(\eta\nu)}, b_2) q^{T}\right\} dq + o(1)\bigg).
\end{multline*}
This is true because $g(\Re T_*^{(\eta\nu)}, b_2) \ne 0$. Performing the Gaussian integration, we obtain
\begin{multline}
\label{almost_end}
F_2(\Lambda) = \left(\frac{2\pi}{n}\right)^{3/2} C_n(X) \cdot \frac{i \exp\{n(\lambda_0^2+b_2^2-2)/2 + \lambda_0(x_1 + x_2)/2\}}{(x_1 - x_2)} \\
\times \bigg(\sum\limits_{\eta \ne \nu} g(\Re T_*^{(\eta\nu)}, b_2) {\det}^{-1/2} \{ -f''(T_*^{(\eta\nu)}, b_2) \} + o(1)\bigg).
\end{multline}
Since
\begin{equation*}
g(\Re T_*^{(\eta\nu)}, b_2) = 2t_\eta^{*} e^{-it_\eta^{*}(x_1 - x_2)},  \quad  \eta \ne \nu,
\end{equation*}
and $C_n$ has the form \eqref{C_n_definition}, we have \eqref{asymptotics_for_F_case_1}.

If $\lambda_0 = 0$, then repeating the above steps, we obtain the formula similar to \eqref{almost_end}. The only difference is that there are two more terms (i.e.\ there are as many terms as stationary points) in the sum. Since $g$ is zero at the points with $t_1= t_2$, we have exactly \eqref{almost_end} and hence the asymptotic equality \eqref{asymptotics_for_F_case_1} is also valid. $\blacksquare$

\medskip
The assertion (i) of the theorem follows immediately from Lemma \ref{lem:1.1}.

\begin{lemma}\label{lem:1.2}
Let all conditions of Theorem \ref{th_1} are hold and $\lambda_0^2 > 4 - 4b_2^2 + \varepsilon$ for some $\varepsilon > 0$. Then $F_2(\Lambda)$ satisfies the asymptotic relation
\begin{enumerate}[(i)]
\item for $\lambda_0 \ne 0$
\begin{equation}
\label{asymptotics_for_F_case_2}
F_2(\Lambda) = \frac{\alpha^2 \exp\{n\widehat{A} + (1 - \alpha)\lambda_0(x_1 + x_2)\}}{(2\alpha-1)^{3/2}(2-\alpha(1-\alpha)(3-2\alpha)\lambda_0^2)^{1/2}}(1 + o(1)),
\end{equation}
where $\alpha$ and $\widehat{A}$ satisfy
\begin{align}
\label{alpha_definition}
\alpha \in (1/2, 1), \qquad &\alpha(1-\alpha)\lambda_0^2+\left(\frac{1-\alpha}{\alpha}\right)^2b_2^2 - 1 = 0, \\
\label{eq:hat A}
&\widehat{A} = f \left(- i\alpha\Lambda_0, \frac{1-\alpha}{\alpha}b_2\right).
\end{align}
\item for $\lambda_0 = 0$
\begin{equation}
\label{asymptotics_for_F_case_2_lambda=0}
F_2(\Lambda) = b_2^n e^{-n/2} \frac{b_2^2}{(b_2^2 - 1)^{3/2}}\left( b_2 + 1 + (-1)^n (b_2 - 1) \right)(1 + o(1)).
\end{equation}
\end{enumerate}
\end{lemma}
\emph{Proof.} Choose $\Im t_1 = \Im t_2 = -\alpha\lambda_0$, $s \in \mathbb{R}$ as the good contour with the stationary point $\left(-i\alpha\lambda_0, -i\alpha\lambda_0, \frac{1-\alpha}{\alpha}b_2\right)$, where $\alpha$ satisfies \eqref{alpha_definition}.
Existence and uniqueness of such $\alpha$ follow from the fact that the l.h.s.\ of \eqref{alpha_definition} is a monotone decreasing function of $\alpha$ whose values at $\alpha = 1/2$ and $\alpha = 1$ have different signs. Everywhere below we assume that $\alpha$ is a solution of \eqref{alpha_definition}.

If $\lambda_0 = 0$, we have two stationary points at the contour --- $(0, 0, \pm 1)$.

Consider the case $\lambda_0 > \varepsilon$. Shifting the variables $t_j \rightarrow t_j - i\alpha\lambda_0$, similarly to \eqref{integral_without_tails} we get
\begin{equation}
\label{integral_without_tails_2}
F_2(\Lambda) = C_n(X) \frac{ie^{(1 - \alpha)\lambda_0(x_1 + x_2)}}{(x_1 - x_2)} \int\limits_{B_R} g(T, s)e^{n f\left(T - i\alpha\Lambda_0, s\right)} dT ds + O(e^{-nR^2/4}).
\end{equation}

It is easy to check that
\begin{align}
\label{f_value}
f \bigg(- i\alpha\Lambda_0, \frac{1-\alpha}{\alpha}b_2\bigg) &= \frac{1}{2}\left(\frac{1-\alpha}{\alpha}\right)^2 b_2^2 + (1-\alpha)\lambda_0^2 + \log{\frac{\alpha}{1-\alpha}} - 1; \\
\notag f'' \left(- i\alpha\Lambda_0, \frac{1-\alpha}{\alpha}b_2\right) &= -\left( \begin{array}{ccc}
1 - (1 - \alpha)^2\lambda_0^2 & \left( \frac{1 - \alpha}{\alpha}\right)^3 b_2^2 & ib_2 \lambda_0\frac{(1 - \alpha)^2}{\alpha} \\
\left( \frac{1 - \alpha}{\alpha}\right)^3 b_2^2 & 1 - (1 - \alpha)^2\lambda_0^2 & ib_2 \lambda_0\frac{(1 - \alpha)^2}{\alpha} \\
ib_2 \lambda_0\frac{(1 - \alpha)^2}{\alpha} & ib_2 \lambda_0\frac{(1 - \alpha)^2}{\alpha} & 1 + b_2^2 \left( \frac{1 - \alpha}{\alpha}\right)^2
\end{array} \right); \\
\label{f_hessian}
\det f'' \bigg(- i\alpha\Lambda_0, \frac{1-\alpha}{\alpha}b_2\bigg) &= -\frac{2\alpha - 1}{\alpha^2} \bigg(\alpha(1 - \alpha)(2\alpha - 1)\lambda_0^2 + 2b_2^2 \left( \frac{1 - \alpha}{\alpha}\right)^2\bigg) < 0; \\
\notag
\det \Re f'' \bigg(- i\alpha\Lambda_0, \frac{1-\alpha}{\alpha}b_2\bigg) &= -\frac{2\alpha - 1}{\alpha^2}\bigg(1 + b_2^2 \left( \frac{1 - \alpha}{\alpha}\right)^2\bigg) \\
\notag
&\times \bigg(\alpha(1 - \alpha)(2\alpha - 1)\lambda_0^2 + b_2^2 \left( \frac{1 - \alpha}{\alpha}\right)^2\bigg) < 0.
\end{align}

In addition, 
\begin{equation*}
\Re f\left(T- i\alpha\Lambda_0, s\right) = h_{\alpha}(t_1, t_2, s, b_2, \lambda_0) + \frac{1}{2}\left(\frac{1-\alpha}{\alpha}\right)^2 b_2^2 + (1-\alpha)\lambda_0^2
\end{equation*}
with $h_\alpha$ of \eqref{h_alpha_definition}. So, similarly to the proof of Lemma \ref{lem:1.1} one can write
\begin{multline}
\label{junction_to_neighborhood_case_2}
F_2(\Lambda) = C_n(X) \frac{ie^{n\widehat{A} + (1 - \alpha)\lambda_0(x_1 + x_2)}}{(x_1 - x_2)} \\
\times \bigg( \int\limits_{V_n} g(T, s)e^{n (f\left(T - i\alpha\Lambda_0, s\right) - \widehat{A})} dT ds + O\left(e^{-C \log^2 n} \right)\bigg),
\end{multline}
where $V_n = U_{n^{-1/2}\log n}\left(\left(0, \frac{1-\alpha}{\alpha}b_2\right)\right)$ and $\widehat{A}$ is defined in \eqref{eq:hat A}.

Repeating the argument of Lemma \ref{lem:1.1}, we get
\begin{multline*}
F_2(\Lambda) = n^{-3/2} C_n(X) \cdot \frac{i \exp\{n\widehat{A} + (1 - \alpha)\lambda_0(x_1 + x_2)\}}{(x_1 - x_2)} (1 + o(1)) \\
\times \int\limits_{[-\log n, \log n]^3} \frac{1}{\sqrt{n}}\bigg(\left<g'\left(0, \frac{1-\alpha}{\alpha}b_2\right), q \right> + \frac{1}{2\sqrt{n}}q g''\left(0, \frac{1-\alpha}{\alpha}b_2\right)q^T \bigg) \\
\times e^{n (f\left(T - i\alpha\Lambda_0, s\right) - \widehat{A})} dq,
\end{multline*}
where $\left< \cdot, \cdot \right>$  denotes the scalar product in $\mathbb{R}^3$. The first integral is zero, because $f$ is a symmetric with respect to $t_1$ and $t_2$ function and $\dfrac{\partial g}{\partial t_1}\left(0, \frac{1-\alpha}{\alpha}b_2\right) = -\dfrac{\partial g}{\partial t_2}\left(0, \frac{1-\alpha}{\alpha}b_2\right) = 1$.
Expanding the exponent into the Taylor series, we obtain
\begin{multline}
\label{almost_end_case_2}
F_2(\Lambda) = n^{-5/2} C_n(X) \cdot \frac{i \exp\{n\widehat{A} + (1 - \alpha)\lambda_0(x_1 + x_2)\}}{(x_1 - x_2)} (1 + o(1)) \\
\times \int\limits_{\mathbb{R}^3} \frac{1}{2}q g''\left(0, \frac{1-\alpha}{\alpha}b_2\right)q^T \exp\left\{-\frac{1}{2} q f''\left(-i\alpha\Lambda_0, \frac{1-\alpha}{\alpha}b_2\right) q^{T}\right\} dq.
\end{multline}

It is easy to see that
\begin{equation*}
\begin{gathered}
g''\left(0, \frac{1-\alpha}{\alpha}b_2\right) = \left( \begin{array}{ccc}
-2ix_1 & i(x_1-x_2) & 0 \\
i(x_1-x_2) & 2ix_2 & 0 \\
0 & 0 & 0
\end{array} \right).
\end{gathered}
\end{equation*}

Computing the integral in \eqref{almost_end_case_2}, we have \eqref{asymptotics_for_F_case_2}.

If $\lambda_0 = 0$, then $\alpha = \frac{b_2}{b_2 + 1}$. As it was mentioned above, in this case there are two stationary points at the contour. The asymptotics of the integral \eqref{integral_without_tails_2} in the neighborhood of the second stationary point (i.e. $(0, 0, -1)$) is computed by the same way as above. \eqref{asymptotics_for_F_case_2} is transformed into \eqref{asymptotics_for_F_case_2_lambda=0}. $\blacksquare$

\medskip
The assertion (ii) follows from \eqref{asymptotics_for_F_case_2} and \eqref{asymptotics_for_F_case_2_lambda=0}.

Now we proceed to the proof of agreement between cases (i) and (ii) of Theorem~\ref{th_1}.
\begin{lemma}
Let all conditions of Theorem \ref{th_1} are hold and $\lambda_0^2 = 4 - 4b_2^2 - \delta_n$, $\delta_n \to 0$. Then $F_2(\Lambda)$ satisfies the asymptotic relation
\begin{equation}
\label{asymptotics_for_F_if_delta}
F_2(\Lambda) = Y_n \exp\{n(2 - 3b_2^2 - \delta_n)/2 + \lambda_0(x_1 + x_2)/2\}(1 + o(1)).
\end{equation}
where
\begin{equation*}
Y_n = \left\{\begin{array}{ll}
n\delta_n^{1/2}, & \text{if }\, \delta_n > 0, n\delta_n^2 \to \infty; \\
Cn^{3/4}, & \text{if }\, n\delta_n^2 \to \text{const}; \\
C(-\delta_n)^{-3/2}, & \text{if }\, \delta_n < 0, n\delta_n^2 \to \infty.
\end{array}\right.
\end{equation*}
\end{lemma}

\emph{Proof.} Consider the case $\delta_n \geq 0$. Choose the same contour as in the proof of Lemma \ref{lem:1.1}. Stationary points are also the same.

Change of the variables $\tau = t_1 + t_2$, $\sigma = t_1 - t_2$ gives us
\begin{equation*}
F_2(\Lambda) = C_n(X) \frac{ie^{\lambda_0(x_1 + x_2)}}{2(x_1 - x_2)}\int\limits_{\mathbb{R}^3} \sigma e^{-i((x_1 + x_2)\tau + (x_1 - x_2)\sigma)/2} e^{n\tilde{f}(\tau, \sigma, s)} d\tau d\sigma ds,
\end{equation*}
where $\tilde{f}(\tau, \sigma, s) = f(T, s)$. Set
\begin{align*}
\tau^* &= -i\lambda_0; \\
\sigma_\eta^* &= t_\eta^* - t_{3 - \eta}^* = (-1)^\eta \sqrt{4 - 4b_2^2 - \lambda_0^2} = (-1)^\eta \delta_n^{1/2}. 
\end{align*}
It is easy to see that
\begin{align*}
\tilde{f}(\tau^*, \sigma_\eta^*, b_2) &= 1 - \frac{3}{2}b_2^2 - \frac{1}{2}\delta_n; \\
\tilde{f}''(\tau^*, \sigma_\eta^*, b_2) &= -\left( \begin{array}{ccc}
b_2^2 + \delta_n/4 & (-1)^\eta i\lambda_0\delta_n^{1/2}/4 & ib_2\lambda_0/2 \\
(-1)^\eta i\lambda_0\delta_n^{1/2}/4 & \delta_n/4 & (-1)^\eta b_2\delta_n^{1/2}/2 \\
ib_2\lambda_0/2 & (-1)^\eta b_2\delta_n^{1/2}/2 & b_2^2 + 1
\end{array} \right); \\
\det \tilde{f}''(\tau^*, \sigma_\eta^*, b_2) &= -\delta_n/4; 
\end{align*}
\begin{align*}
\frac{\partial^3 \tilde{f}}{\partial \sigma^3}(\tau^*, \sigma_\eta^*, b_2) &= \frac{(-1)^{\eta + 1}}{4} \delta_n^{1/2}(\delta_n - 3); \\
\frac{\partial^4 \tilde{f}}{\partial \sigma^4}(\tau^*, \sigma_\eta^*, b_2) &= -\frac{3}{4}.
\end{align*}

Let us choose $V_n$ as a union of the products of the neighborhoods of $\tau^*$, $\sigma_1^*$, $b_2$ and $\tau^*$, $\sigma_2^*$, $b_2$ such that the radii of the neighborhoods corresponding to $\tau$ and $s$ are equal to $\log n/\sqrt{n}$, whereas the radius of the neighborhood corresponding to $\sigma$ is equal to $\log n/\sqrt{n\delta_n}$, if $n\delta_n^2 \to \infty$, and to $\log n/n^{1/4}$ otherwise. Similarly to the proof of Lemma \ref{lem:1.1} it can be proved that for $\left(\tau, \sigma, s\right) \notin V_n$ and sufficiently big $n$
\begin{equation}
\label{SDM_key_inequality}
\Re \tilde{f}(\tau^*, \sigma_\eta^*, b_2) - \Re \tilde{f}(\tau, \sigma, s) \geq C\frac{\log^2 n}{n},
\end{equation}

Let $n\delta_n^2 \to \infty$. Then by the same way as before, with the only one distinction that the change of the variable $\sigma$ is $\sigma \rightarrow (n\delta_n)^{-1/2}\sigma + \sigma_\eta^*$, we obtain
\begin{multline}\label{eq:after Taylor expanding (big delta)}
F_2(\Lambda) = n^{-3/2} C_n(X) \cdot \frac{i \exp\{n(2 - 3b_2^2 - \delta_n)/2 + \lambda_0(x_1 + x_2)/2\}}{2(x_1 - x_2)} (1 + o(1)) \\
\times \sum\limits_{\eta =1}^2 \int\limits_{[-\log n, \log n]^3} \left((-1)^\eta + i(x_1 - x_2)\delta_n^{1/2}/2 + \frac{\sigma}{\delta_n\sqrt{n}} \right) \\
\times e^{\frac{1}{2} A_{\tilde{f}}^{(\eta)}(\tau, \sigma, s)} d\tau d\sigma ds,
\end{multline}
where $A_{\tilde{f}}^{(\eta)}$ is a quadratic form, defined by the matrix, which is obtained from $\tilde{f}''(\tau^*, \sigma_\eta^*, b_2)$ by dividing by $\delta_n^{1/2}$ of all numbers in the second line and the second column, i.e.
\begin{equation*}
A_{\tilde{f}}^{(\eta)}(\tau, \sigma, s) = (\tau, \delta_n^{-1/2}\sigma, s)\tilde{f}''(\tau^*, \sigma_\eta^*, b_2)(\tau, \delta_n^{-1/2}\sigma, s)^{T}.
\end{equation*}
Therefore we have \eqref{asymptotics_for_F_if_delta}.

Now let $n\delta_n^2 \to 0$. Then, changing the variables $\sigma^2 = \tilde{\sigma}$, we get
\begin{equation*}
F_2(\Lambda) = C_n(X) \frac{e^{\lambda_0(x_1 + x_2)}}{2(x_1 - x_2)}\int\limits_0^{+\infty} d\tilde{\sigma} \int\limits_{\mathbb{R}^2} \sin((x_1 - x_2)\sqrt{\tilde{\sigma}}/2)e^{-i(x_1 + x_2)\tau/2} e^{n\tilde{f}(\tau, \sqrt{\tilde{\sigma}}, s)} d\tau ds.
\end{equation*}
Let $\tilde{V}_n$ be a product of the $\frac{\log n}{\sqrt{n}}\text{-neighborhoods}$ of 0 and $b_2$. Then we can shift the variable $\tau \rightarrow \tau - i\lambda_0$ and in view of \eqref{SDM_key_inequality} restrict the integration domain  by $\left[0, \frac{\log^2 n}{\sqrt{n}}\right] \times \tilde{V}_n$
\begin{multline*}
F_2(\Lambda) = C_n(X) \frac{e^{\lambda_0(x_1 + x_2)/2}}{2(x_1 - x_2)} \int\limits_0^{\frac{\log^2 n}{\sqrt{n}}}(1 + o(1)) d\tilde{\sigma} \\
\times \int\limits_{\tilde{V}_n} \sin((x_1 - x_2)\sqrt{\tilde{\sigma}}/2)e^{-i(x_1 + x_2)\tau/2} e^{n\tilde{f}(\tau - i\lambda_0, \sqrt{\tilde{\sigma}}, s)} d\tau ds.
\end{multline*}
Expanding $\tilde{f}$ and sin by the Taylor formula near $(-i\lambda_0, 0, b_2)$ and 0 respectively and changing the variables $\tau \rightarrow n^{-1/2}\tau$, $\tilde{\sigma} \rightarrow n^{-1/2}\tilde{\sigma}$, $s \rightarrow n^{-1/2}s + b_2$, we have
\begin{multline}\label{eq:after Taylor expanding (small delta)}
F_2(\Lambda) = n^{-3/2}C_n(X) \frac{\exp\{n(2 - 3b_2^2 - \delta_n)/2 + \lambda_0(x_1 + x_2)/2\}}{2(x_1 - x_2)} \\
\times \int\limits_0^{\log^2 n} (1 + o(1))d\tilde{\sigma} \int\limits_{[-\log n,\, \log n]^2} (x_1 - x_2)\frac{\sqrt{\tilde{\sigma}}}{2\sqrt[4]{n}} e^{\frac{1}{2} B_{\tilde{f}}(\tau, \tilde{\sigma}, s)} d\tau ds,
\end{multline}
where $B_{\tilde{f}}$ is a quadratic form defined by the matrix
\begin{equation*}
\left( \begin{array}{ccc}
\frac{\partial^2 \tilde{f}}{\partial \tau^2} & \frac{1}{2} \frac{\partial^3 \tilde{f}}{\partial \tau\partial\sigma^2} & \frac{\partial^2 \tilde{f}}{\partial \tau\partial s} \\
\frac{1}{2} \frac{\partial^3 \tilde{f}}{\partial \tau\partial\sigma^2} & \frac{1}{12} \frac{\partial^4 \tilde{f}}{\partial \sigma^4} & \frac{1}{2} \frac{\partial^3 \tilde{f}}{\partial \sigma^2 \partial s} \\
\frac{\partial^2 \tilde{f}}{\partial \tau\partial s} & \frac{1}{2} \frac{\partial^3 \tilde{f}}{\partial \sigma^2 \partial s} & \frac{\partial^2 \tilde{f}}{\partial s^2}
\end{array} \right)(-i\lambda_0, 0, b_2) = -\left( \begin{array}{ccc}
b_2^2 & i\lambda_0/8 & ib_2\lambda_0/2 \\
i\lambda_0/8 & 1/16 & b_2/4 \\
ib_2\lambda_0/2 & b_2/4 & 1 + b_2^2
\end{array} \right) + o(1)P,
\end{equation*}
where all entries of the matrix $P$ are units. Performing the Gaussian integration, we obtain
\begin{multline*}
F_2(\Lambda) = \frac{2\pi}{n^{3/2}} \cdot \frac{C_n(X)}{4\sqrt[4]{n}} \cdot \exp\{n(2 - 3b_2^2 - \delta_n)/2 + \lambda_0(x_1 + x_2)/2\} \\
\int\limits_0^{+\infty} \sqrt{\tilde{\sigma}} \exp\left\{-\frac{\tilde{\sigma}^2}{32}\right\} d\tilde{\sigma} (1 + o(1))
\end{multline*}
that imply \eqref{asymptotics_for_F_if_delta}.

If $n\delta_n^2 \to \text{const}$, then there is a certain third power polynomial instead of $B_{\tilde{f}}$ in the last exponent in \eqref{eq:after Taylor expanding (small delta)}. Thus, the asymptotics of $F_2(\Lambda)$ differs from \eqref{asymptotics_for_F_if_delta} only by multiplicative $n$-independent constant, with is absorbed by $C$ in $Y_n$. For negative $\delta_n$, if $n\delta_n^2 \to 0$, all the changes in equations appear in multiplication by factors which are equal to $1 + o(1)$, so \eqref{asymptotics_for_F_if_delta} is unchanged
. If $n\delta_n^2 \to \infty$, then the combination of the argument of the case $\delta_n \geq 0$ and of Lemma \ref{lem:1.2} causes some changes in \eqref{eq:after Taylor expanding (big delta)} and implies $Y_n = C(-\delta_n)^{-3/2}$ in \eqref{asymptotics_for_F_if_delta}. $\blacksquare$


\section{Proof of Theorem \ref{th_3}}\label{sec:th3}

As in the case of Theorem \ref{th_1}, the proof of Theorem \ref{th_3} is based on the application of the steepest descent method to the integral representation of $F_{2m}$ obtained in Section \ref{sec:int_repres}. For this end a ``good contour'' and stationary points of $f_{2m}$, defined in \eqref{f_2m_definition}, have to be chosen. We start from the choice of stationary points.

If $p = n$ and hence $b_l = \tilde{b}_{2k} = 0$, $l > 1$, the proper stationary points are
\begin{equation*} 
t_j = t_j^* = 
\begin{cases}
\phantom{-}t^* \\
-\overline{t^*}
\end{cases},\, j = \overline{1,2m}, \quad R_l = 0,
\end{equation*}
where
\begin{equation*}
t^* = \frac{1}{2}\left(-i\lambda_0 + \sqrt{4 - \lambda_0^2}\right).
\end{equation*}

Set 
\begin{gather}
\label{eq:T tilde def}
\widetilde{T} = \diag\{t_j^*\}_{j = 1}^{2m}, \\
\notag
b = (b_2, \ldots, b_{2m}, \tilde{b}_4, \tilde{b}_6, \ldots, \tilde{b}_{2m}).
\end{gather}
Since
\begin{multline*} 
\left.\left(A_{2m}(T, R)f_{2m}'(T, R)\right)'\right|_{\substack{T = \widetilde{T}\\R = 0\\b = 0}} \\
= (-1)^{m - 1}\diag\{1 + \overline{t_1^*}^2, \ldots, 1 + \overline{t_{2m}^*}^2, e_1, \ldots, e_{\binom{4m}{2m} - 4m^2 - 1}\},\quad e_j = 1 \text{ or } 2
\end{multline*}
is nondegenerate for $\lambda_0 \in (-2, 2)$, for sufficiently small $b$ there exists the unique solution
\begin{equation}\label{eq:system solution}
T = T(b),\, R = R(b)
\end{equation}
of the equation 
\begin{equation}\label{eq:stationary system}
A_{2m}(T, R)f_{2m}'(T, R) = 0
\end{equation}
such that $T(0) = \widetilde{T}$, $R(0) = 0$, and the solution continuously depends on $b$ and $\lambda_0$. When $p \to \infty$, \eqref{eq:a_l asympt} and \eqref{eq:b_l definition} yield $b \to 0$. Therefore, the solutions \eqref{eq:system solution} are the required stationary points. 

\begin{lemma} 
The solution \eqref{eq:system solution} also has the following properties
\begin{itemize}
\item[(1)] $T_{j_1 j_1}(b) = T_{j_2 j_2}(b)$, $T_{k_1 k_1}(b) = T_{k_2 k_2}(b)$ for all $j_1, j_2 \in I_+$, $k_1, k_2 \in I_-$, where $I_+ = \{j,\, t_j^* = t^*\}$, $I_- = \{j,\, t_j^* = -\overline{t^*}\}$;
\item[(2)] $(R_l)_{\alpha\beta}(b) = 0$, $l = \overline{2, 2m}$, $\alpha \ne \beta$.
\end{itemize}
\end{lemma}

\emph{Proof.} Let
\begin{align*}
\pi T = \diag\{t_{\pi(j)}\}, \quad \pi R = (\pi R_2, \ldots, \pi R_{2m}), \quad \pi \in S_{2m},
\end{align*}
where $\pi R_l$ is the such matrix that $(\pi R_l)_{\alpha\beta} = (R_l)_{\alpha_\pi \beta_\pi}$. Then $\forall \pi \in S_{2m}$ $f_{2m}(\pi T, \pi R) = f_{2m}(T, R)$. So, it is sufficiently to proof the lemma for those stationary points, for which $t_1^* = \ldots = t_k^* = -\overline{t_{k + 1}^*} = \ldots = -\overline{t_{2m}^*} = t^*$.

We are going to prove that there exists the solution of \eqref{eq:stationary system} that satisfies conditions (1)-(2) and $T(0) = \widetilde{T}$, $R(0) = 0$. It is equivalent to existence of the solution of the system
\begin{equation}
\begin{split}
\label{eq:system 2}
A_{2m}(T, R)\frac{\partial}{\partial t_j}f_{2m}(T, R) &= 0, \quad j = 1, 2m; \\
A_{2m}(T, R)\frac{\partial}{\partial (R_l)_{\alpha\alpha}}f_{2m}(T, R) &= 0, \quad \alpha \in I_{2m,l},\, l = \overline{2,2m},
\end{split}
\end{equation}
where $T$ and $R$ satisfy (1)-(2), with respect to the variables $t_1$, $t_{2m}$, $(R_l)_{\alpha\alpha}$. Since the derivative of the l.h.s. of the system at $T = \widetilde{T}$, $R = 0$, $b = 0$ is nondegenerate, there exists the solution of it. In view of uniqueness of the solution of \eqref{eq:stationary system}, the solution of \eqref{eq:system 2} coincides with \eqref{eq:system solution}. $\blacksquare$

The next step is the choice of the ``contour'' (in this case it is a $d_{2m}$-dimensional manifold, $d_{2m} = \left(\binom{4m}{2m} - 4m^2 - 1 + 2m\right)$). For each variable we consider some contour and the required manifold $\hat{M}_{2m}$ will be the product of these contours. Fix some variable. Order the corresponding components of the stationary points by increasing of the real part (if real parts are equal, order by increasing of the imaginary part). Then the contour is a polygonal chain that connect points by the order described above. Infinite segments of the polygonal chain are parallel to the real axis and directed from the first point to the left and from the last point to the right.

The Cauchy theorem and \eqref{final_integral_representation_m>1} imply
\begin{equation*}
F_{2m}(\Lambda) = C_n^{(2m)}(X) \frac{i^{2m^2 - m}\exp\Big\{\lambda_0\sum\limits_{j = 1}^{2m} x_j\Big\}}{\Delta(X)} \int\limits_{\hat{M}_{2m}} g_{2m}(T) e^{nf_{2m}(T, R)} dT dR,
\end{equation*}
where
\begin{equation*}
g_{2m}(T) = \Delta(T) \exp\bigg\{-i\sum\limits_{j=1}^{2m} x_jt_j\bigg\}.
\end{equation*}
Moreover,
\begin{equation*}
F_{2m}(\Lambda) = C_n^{(2m)}(X) \frac{i^{2m^2 - m}\exp\Big\{\lambda_0\sum\limits_{j = 1}^{2m} x_j\Big\}}{\Delta(X)} \bigg( \int\limits_{\hat{M}_{2m}^N} g_{2m}(T) e^{nf_{2m}(T, R)} dT dR + r(n, N)\bigg),
\end{equation*}
where
\begin{equation*}
\begin{gathered}
\hat{M}_{2m}^N = \{\zeta \in \hat{M}_{2m} | \left\|\Re \zeta\right\|_{\infty} \le N\}, \\
|r(n, N)| < Ce^{-nN^2/4},\, N \to \infty,
\end{gathered}
\end{equation*}
with $\left\| y \right\|_\infty = \max\limits_j |y_j|$.

Let $\varepsilon > 0$ be an arbitrary positive number. Then, since $(T(b), R(b)) \underset{n \to \infty} {\longrightarrow} (\widetilde{T}, 0)$, for sufficiently big $n$ and for every $(T, R) \in \hat{M}_{2m}^N$ we have
\begin{equation*}
\left|\Re f_{2m}^0\bigg(\Re T - \frac{i}{2}\Lambda_0, \Re R\bigg) - \Re f_{2m}^0(T, R)\right| < \varepsilon,
\end{equation*}
where $f_{2m}^0(T, R) = f_{2m}(T, R)|_{b = 0}$. Also, $f_{2m}(T, R) \rightrightarrows f_{2m}^0(T, R)$, $n \to \infty$, $(T, R) \in K$ for any compact set $K$. Hence, for sufficiently big $n$
\begin{equation*}
| \Re f_{2m}(T, R) - \Re f_{2m}^0(T, R)| < \varepsilon, \quad (T, R) \in \hat{M}_{2m}^N.
\end{equation*}

Consider the point $(T^0, R^0) \in \hat{M}_{2m}^N$ such that $\Re T^0 = \Re \widetilde{T}$, $\Re R^0 = 0$. Then $\Re f_{2m}(T^0, R^0) > \Re f_{2m}^0(\widetilde{T}, 0) - 2\varepsilon$. Thus,
\begin{equation*}
\max_{(T, R) \in \hat{M}_{2m}^N} \Re f_{2m}(T, R) > \Re f_{2m}^0(\widetilde{T}, 0) - 2\varepsilon = \max_{\substack{\Im T = \Lambda_0/2\\|\Re t_j| \le N}} \Re f_{2m}^0 (T, 0) - 2\varepsilon.
\end{equation*}
Therefore, if $\Re f_{2m}(T^1, R^1) = \max\limits_{(T, R) \in \hat{M}_{2m}^N} \Re f_{2m}(T, R)$, then
\begin{multline*}
(T^1, R^1) \in \{(T, R) \in \hat{M}_{2m}^N |\, \Re f_{2m}(T, R) > \Re f_{2m}^0(\widetilde{T}, 0) - 2\varepsilon\} \\
\subset \{(T, R) \in \hat{M}_{2m}^N |\, \Re f_{2m}^0(\Re T - \frac{i}{2}\Lambda_0, \Re R) > \Re f_{2m}^0(\widetilde{T}, 0) - 4\varepsilon\}.
\end{multline*}
So, it is evident, that $(T^1, R^1) \to (\widetilde{T}, 0)$, $n \to \infty$ for certain $\widetilde{T}$ of the form \eqref{eq:T tilde def}.

Let $V_n(T(b), R(b))$ be the neighborhood of the stationary point $(T(b), R(b))$, which contains the corresponding maximum point of $\Re f_{2m}$ with its $\frac{\log n}{\sqrt{n}}$-neighborhood, and $\operatorname{diam} V_n(T(b), R(b)) \to 0$. It can be assumed that the union of these neighborhoods is invariant for map $(T, R) \rightarrow (\pi T, \pi R)$ for every $\pi \in S_{2m}$. Then, by the same reasons as in the proof of Theorem \ref{th_1}, we can restrict the integration domain by the union of the neighborhoods $V_n$. Shifting the variables $T \rightarrow T + \Im T(b)$, $R \rightarrow R + \Im R(b)$ in each neighborhood and expanding $g_{2m}$ by the Taylor formula, we get
\begin{multline}
\label{eq:junction to the neighborhood, m > 1}
F_{2m}(\Lambda) = C_n^{(2m)}(X) \frac{i^{2m^2 - m}\exp\Big\{\lambda_0\sum\limits_{j = 1}^{2m} x_j\Big\}}{\Delta(X)}(1 + o(1)) \\
\times \sum e^{nf_{2m}(T(b), R(b))}\int\limits_{\hat{V}_n(T(b), R(b))} \bigg(\sum\limits_{|\alpha| \le 2m(m - 1)}n^{-|\alpha|/2}D^\alpha g_{2m}(T(b))t^\alpha \\
+ r_{g_{2m}}^{(2m(m - 1))}(T)\bigg) e^{nf_{2m}(T + \Im T(b), R + \Im R(b))} dT dR,
\end{multline}
where the summation is over all stationary points under consideration and
\begin{equation*}
\begin{gathered}
\hat{V}_n(T(b), R(b)) = V_n(T(b), R(b)) - \Im(T(b), R(b)); \\
|r_{g_{2m}}^{(2m(m - 1))}(T)| \le C\sum\limits_{j} |t_j|^{2m(m - 1) + 1}.
\end{gathered}
\end{equation*}
The number of terms of the Taylor series in \eqref{eq:junction to the neighborhood, m > 1} is the minimal number that allows us to obtain nonzero asymptotics.

Fix some stationary point $(T(b),R(b))$ and some multi-index $\alpha$. Let $\beta \le \alpha$ be a multi-index with $\beta_{j_1} = \beta_{j_2}$ for some $j_1 \ne j_2$, $j_1, j_2 \in I_+$ or $j_1, j_2 \in I_-$, where $\beta \le \alpha \Leftrightarrow \forall j\ \beta_j \le \alpha_j$. Then
\begin{multline*}
\int\limits_{\hat{V}_n(T(b), R(b))} \bigg(D^{\alpha - \beta}\Delta(T)D^\beta \exp\bigg\{-i\sum\limits_{j = 1}^{2m}x_j t_j\bigg\}\bigg|_{T = T(b)} t^\alpha \\
+ D^{\hat{\alpha} - \beta}\Delta(T)D^\beta \exp\bigg\{-i\sum\limits_{j = 1}^{2m}x_j t_j\bigg\}\bigg|_{T = T(b)} t^{\hat{\alpha}}\bigg) e^{nf_{2m}(T + \Im T(b), R + \Im R(b))} dTdR = 0,
\end{multline*}
where $\hat{\alpha}$ is the multi-index, which is obtained by swapping $\alpha_{j_1}$ and $\alpha_{j_2}$ in $\alpha$. Hence, in the sum in \eqref{eq:junction to the neighborhood, m > 1} only the summands with $|\alpha| = 2m(m - 1)$ remain.

Changing the variables $T \rightarrow n^{-1/2}T$, $R \rightarrow n^{-1/2} R$, we get
\begin{multline}
\label{eq:Taylor expansion in the integral (case p->inf)}
F_{2m}(\Lambda) = n^{-d_{2m}/2} C_n^{(2m)}(X) \frac{i^{2m^2 - m}\exp\Big\{\lambda_0\sum\limits_{j = 1}^{2m} x_j\Big\}}{\Delta(X)}(1 + o(1)) \sum\limits \bigg[ e^{nf_{2m}(T(b), R(b))} \\
\times \int\limits_{\mathcal{V}_n(T(b), R(b))} \bigg(n^{-m(m - 1)}\sum\limits_{|\alpha| = 2m(m - 1)}D^\alpha g_{2m}(T(b))t^\alpha + r_{g_{2m}}^{(2m(m - 1))}\left(\frac{1}{\sqrt{n}}T\right)\bigg) \\
\times \exp\left\{\frac{1}{2} q f_{2m}''(T(b), R(b)) q^T\right\}\left(1 + \frac{r_6(q)}{\sqrt{n}}\right) dq\bigg],
\end{multline}
where $q$ is a vector which consists of all integration variables, $dq = dTdR$, $\mathcal{V}_n(T(b), R(b)) = \sqrt{n}\hat{V}_n(T(b), R(b))$ and $|r_6(q)| \le C \sum\limits_{j} |q_j|^3$.

\eqref{eq:system 2} implies that, as $n \to \infty$,
\begin{align*}
(R_l)_{\alpha\alpha} &= o(b_2), \quad l = \overline{3, 2m}; \\
(R_2)_{\alpha\alpha} &= \frac{b_2}{T_{\alpha_1 \alpha_1}(0)T_{\alpha_2 \alpha_2}(0)} + o(b_2); \\
T_{jj}(b) &= T_{jj}(0) - \frac{(2m - \chi(j))T_{jj}(0) + \chi(j) - 1}{T_{jj}(0)^3 + T_{jj}(0)^5}b_2^2 + o(b_2^2),
\end{align*}
where
\begin{equation*}
\chi(j) =
\begin{cases}
\#I_+, \quad \text{if } j \in I_+; \\
\#I_-, \quad \text{if } j \in I_-.
\end{cases}
\end{equation*}
Therefore,
\begin{equation*}
\Re f_{2m}(T(b), R(b)) = (-1)^m + \chi(j)(2m - \chi(j))\cdot \frac{\lambda_0^2(4 - \lambda_0^2)}{2}b_2^2 + o(b_2^2).
\end{equation*}
Thus, the value of the $\Re f_{2m}$ at the stationary points of the form \eqref{eq:system solution} with $\#I_+ = m$ is greater than that at the other stationary points of such a form for $\lambda_0 \in (-2, 2)\backslash \{0\}$. For $\lambda_0 = 0$ the values of the $\Re f_{2m}$ at the stationary points are equal, because \eqref{eq:system solution} continuously depends on $\lambda_0$. This yields that the sum in \eqref{eq:Taylor expansion in the integral (case p->inf)} may be restricted to the sum only over the stationary points with $\#I_+ = m$ (for $\lambda_0 = 0$ the other summands have the less order of $n$). We have
\begin{multline*}
F_{2m}(\Lambda) = n^{-d_{2m}/2 - m(m - 1)} C_n^{(2m)}(X) \frac{i^{2m^2 - m}\exp\Big\{\lambda_0\sum\limits_{j = 1}^{2m} x_j\Big\}}{\Delta(X)}(1 + o(1)) \\
\times \bigg(\sum\limits_{\#I_+ = m} e^{nf_{2m}(T(b), R(b))} \int\limits_{\mathbb{R}^{d_{2m}}} \sum\limits_{|\alpha| = 2m(m - 1)}D^\alpha g_{2m}(T(b))t^\alpha \\
\times \exp\left\{\frac{1}{2} q f_{2m}''(T(b), R(b)) q^T\right\} dq\bigg).
\end{multline*}
Consider the term with $I_+ = \{1, 2, \ldots, m\}$. The Gaussian integration gives us
\begin{multline*}
C n^{m^2} i^{3m^2 - 2m}\frac{\Delta(x_1, \ldots, x_m)\Delta(x_{m + 1}, \ldots, x_{2m})}{\Delta(X)}\exp\bigg\{\frac{mn}{2}(\lambda_0^2 - 2) + \frac{\lambda_0}{2}\sum\limits_{j = 1}^{2m} x_j\bigg\}\\
\times \exp\bigg\{-\frac{i\sqrt{4 - \lambda_0^2}}{2}\sum\limits_{j = 1}^m(x_j - x_{m + j})\bigg\}(1 + o(1)) 
= C n^{m^2} (-1)^{m(m + 1)/2} \\
\times \frac{\exp\Big\{-\frac{i\sqrt{4 - \lambda_0^2}}{2}\sum\limits_{j = 1}^m(x_j - x_{m + j})\Big\}}{i^{m}\prod\limits_{j,k = 1}^m (x_j - x_{m + k})} \exp\bigg\{\frac{mn}{2}(\lambda_0^2 - 2) + \frac{\lambda_0}{2}\sum\limits_{j = 1}^{2m} x_j\bigg\}(1 + o(1)),
\end{multline*}
where $C$ is some real $n$-independent constant.

On the other hand,
\begin{equation*}
\hat{S}_{2m}(X) = \frac{\det \left\{ \dfrac{e^{i\pi\rho_{sc}(\lambda_0)(x_j - x_{m + k})} - e^{-i\pi\rho_{sc}(\lambda_0)(x_j - x_{m + k})}}{(x_j - x_{m + k})}\right\}_{j,k = 1}^m}{(2i\pi\rho_{sc}(\lambda_0))^m \Delta(x_1, \ldots, x_m)\Delta(x_{m + 1}, \ldots, x_{2m})},
\end{equation*}
where $\hat{S}_{2m}$ is defined in \eqref{eq:hat S_2m definition} and $\rho_{sc}$ is defined in \eqref{eq:semicircle}. The determinant in the l.h.s. is the sum of $\exp\Big\{i\pi\rho_{sc}(\lambda_0)\sum\limits_{j = 1}^m \varepsilon_j x_j\Big\}$ over all collections $\{\varepsilon_j\}$, which consist of $m$ elements $+1$ and $m$ elements $-1$, with certain coefficients. Since (see \cite[Problem 7.3]{Po-Sz:76})
\begin{equation*}
(-1)^{m(m - 1)/2}\frac{\prod\limits_{j < k} (u_j - u_k)(v_j - v_k)}{\prod\limits_{j,k = 1}^m (u_j - v_k)} = \det \left\{\frac{1}{u_j - v_k}\right\}_{j,k = 1}^m,
\end{equation*}
the coefficient under $\exp\Big\{-i\pi\rho_{sc}(\lambda_0)\sum\limits_{j = 1}^m (x_j - x_{m + j})\Big\}$ is
\begin{equation*}
\frac{\det \left\{ \dfrac{1}{(x_{m + j} - x_{k})}\right\}_{j,k = 1}^m}{(2i\pi\rho_{sc}(\lambda_0))^m \Delta(x_1, \ldots, x_m)\Delta(x_{m + 1}, \ldots, x_{2m})} = \frac{(-1)^{m(m + 1)/2}}{(2i\pi\rho_{sc}(\lambda_0))^m \prod\limits_{j,k = 1}^m (x_j - x_{m + k})}.
\end{equation*}
The other coefficients can be computed by the same way. Therefore,
\begin{equation*}
F_{2m}(\Lambda) = Cn^{m^2} \exp\bigg\{\frac{mn}{2}(\lambda_0^2 - 2) + \frac{\lambda_0}{2}\sum\limits_{j = 1}^{2m} x_j\bigg\}\hat{S}_{2m}(X)(1 + o(1)).
\end{equation*}
The assertion of the theorem follows.

\section{Proof of Theorem \ref{th_2}}\label{sec:th2}

In this section we consider the case $p \to \infty$. As in the proof of Lemma \ref{lem:1.2}, the good contour is $\Im t_1 = \Im t_2 = -\alpha\lambda_0$, $s \in \mathbb{R}$ with the stationary point $\left(-i\alpha\lambda_0, -i\alpha\lambda_0, \frac{1-\alpha}{\alpha}b_2\right)$, where $\alpha$ satisfies \eqref{alpha_definition}.

Set $\beta = 2\alpha - 1$. Then \eqref{alpha_definition} transforms to $b_2 = \beta(1 + \beta)(1 - \beta)^{-1}$, and hence $\beta = \sqrt{\frac{2}{p}}(1 + o(1))$. Substitution of $\lambda_0 = \pm 2$, $\alpha = \frac{1 + \beta}{2}$ and $b_2 = \beta(1 + \beta)(1 - \beta)^{-1}$ in \eqref{f_value}-\eqref{f_hessian} yields
\begin{align*}
f'' \bigg(- i\alpha\Lambda_0, \frac{1-\alpha}{\alpha}b_2\bigg) &= -\left( \begin{array}{ccc}
\beta(2 + \beta) & \beta^2(1 - \beta)(1 + \beta)^{-1} & \pm i\beta(1 - \beta) \\
\beta^2(1 - \beta)(1 + \beta)^{-1} & \beta(2 + \beta) & \pm i\beta(1 - \beta) \\
\pm i\beta(1 - \beta) & \pm i\beta(1 - \beta) & 1 + \beta^2
\end{array} \right); \\
\det f'' \bigg(- i\alpha\Lambda_0, \frac{1-\alpha}{\alpha}b_2\bigg) &= -\frac{4\beta}{(1 + \beta)^2}\cdot (2 - (1 + \beta)(1 - \beta)(2 - \beta)) = -\frac{4\beta^2(1 + 2\beta - \beta^2)}{(1 + \beta)^2}.
\end{align*}
In addition,
\begin{align*}
\frac{\partial^3 f}{\partial t_j^3}\left(-i\alpha\Lambda_0, \frac{1-\alpha}{\alpha}b_2\right) &= -2i\operatorname{sign}(\lambda_0)(1 - \beta)^3.
\end{align*}
If (i) $\frac{n^{2/3}}{p} \to \infty$, then $V_n$ is chosen as a product of the neighborhoods of $0$, $0$, and $\frac{1 - \alpha}{\alpha} b_2$ such that the radius of the corresponding to $t_1$ and $t_2$ neighborhoods is $\frac{\log n}{\sqrt{n\beta}}$, but the radius of the corresponding to $s$ neighborhood is $\frac{\log n}{\sqrt{n}}$. We have \eqref{junction_to_neighborhood_case_2}.

Change of variables $T \rightarrow (n\beta)^{-1/2}T$, $s \rightarrow n^{-1/2}s + \frac{1 - \alpha}{\alpha} b_2$ and repeating of the argument of the proof of Lemma \ref{lem:1.2} yield
\begin{equation*}
F_2(2I + n^{-2/3}X) = Cp \exp\{n\widehat{A} + n^{1/3}\lambda_0(x_1 + x_2)/2\}(1 + o(1)),
\end{equation*}
where $C$ is some absolute constant and $\widehat{A} = f \left(- i\alpha\Lambda_0, \frac{1-\alpha}{\alpha}b_2\right)$.\\
The assertion (i) follows.

Let now (ii) $\frac{n^{2/3}}{p} \to c$. Chose $V_n$ the same as in the case (i), but the radius of the neighborhoods corresponding to $t_1$ and $t_2$ is $\frac{\log n}{\sqrt[3]{n}}$. Then \eqref{junction_to_neighborhood_case_2} is also valid. \\
The Cauchy theorem implies
\begin{multline*}
F_2(2I + n^{-2/3}X) = C_n(X) \frac{ie^{n\widehat{A} + 2n^{1/3}(1 - \alpha)(x_1 + x_2)}}{n^{1/3}(x_1 - x_2)} \bigg( \int\limits_{W_n} g(T, s)e^{n (f\left(T - 2i\alpha I, s\right) - \widehat{A})} dT ds \\
+ O\left(e^{-C \log^2 n} \right)\bigg),
\end{multline*}
where the integration domain over $s$ is not changed, but the ones over $t_j$ become $\{ |z| \leq n^{-1/3}\log n\ |\ \arg z = -\pi/6 \text{ or } \arg z = -5\pi/6 \}$.\\
Changing the variables $T \rightarrow n^{-1/3}T$, $s \rightarrow n^{-1/2}s + \frac{1 - \alpha}{\alpha} b_2$, we obtain
\begin{multline*}
F_2(2I + n^{-2/3}X) = n^{-5/3} C_n(X) \cdot \frac{i \exp\{n\widehat{A} + 2n^{1/3}(1 - \alpha)(x_1 + x_2)\}}{(x_1 - x_2)} (1 + o(1)) \\
\times \int\limits_{\mathbb{R}} e^{\frac{1 + \beta^2}{2} s^2} ds \int\limits_{\gamma \times \gamma} \left( t_1 - t_2 \right) \exp\bigg\{ -\sum\limits_{j = 1}^2 \left(\frac{i}{3}(1 - \beta)^3 t_j^3 + \frac{(2 + \beta)\sqrt{c}}{\sqrt{2}} t_j^2 + i x_j t_j\right)\bigg\} dT
\end{multline*}
where
\begin{equation*}
\gamma = \{\arg z = -\pi/6 \text{ or } \arg z = -5\pi/6\}.
\end{equation*}
Change of the variables $\tau_j = t_j + i\sqrt{2c}$ and using of the Cauchy theorem gives us
\begin{multline*}
F_2(2I + n^{-2/3}X) = Cn^{2/3} \cdot \frac{i \exp\{n\widehat{A} + (2n^{1/3}(1 - \alpha) + \sqrt{2c})(x_1 + x_2)\}}{(x_1 - x_2)} (1 + o(1)) \\
\times \int\limits_{\gamma \times \gamma} \left( \tau_1 - \tau_2 \right) \exp\bigg\{ -\sum\limits_{j = 1}^2 \left(\frac{i}{3}\tau_j^3 + i (x_j + 2c) \tau_j\right)\bigg\} d\tau_1 d\tau_2,
\end{multline*}
where $C$ is some absolute constant. Taking into account \eqref{eq:Airy kernel} and \eqref{C_n_definition}, we get
\begin{multline*}
F_2(2I + n^{-2/3}X) = Cn^{2/3} \exp\{n\widehat{A} + (2n^{1/3}(1 - \alpha) + \sqrt{2c})(x_1 + x_2)\} \\
\times \mathbb{A}(x_1 + 2c, x_2 + 2c)(1 + o(1)).
\end{multline*}
which completes the proof of the assertion (ii).

\section{Appendix}
\subsection{Grassmann variables}\label{sec:grassmanns}

Consider the set of formal variables $\{\psi_j,\, \overline{\psi}_j\}_{j=1}^n$ which satisfy the anticommutation relations
\begin{equation*}
\psi_j \psi_k + \psi_k \psi_j = \overline{\psi}_j \psi_k + \psi_k \overline{\psi}_j = \overline{\psi}_j \overline{\psi}_k + \overline{\psi}_k \overline{\psi}_j = 0.
\end{equation*}
This set generates a graded algebra $\mathcal{A}$, which is called the Grassmann algebra. Taking into account that $\psi_j^2 = \overline{\psi}_j^2 = 0$, we have that all elements of $\mathcal{A}$ are polynomials of $\{\psi_j,\, \overline{\psi}_j\}_{j=1}^n$. We can also define functions of Grassmann variables. Let $\chi$ be an element of $\mathcal{A}$ and $f$ be any analytical function. By $f(\chi)$ we mean the element of $\mathcal{A}$ obtained by substituting $\chi - z_0$ in the Taylor series of $f$ near $z_0$, where $z_0$ is a free term of $\chi$. Since $\chi - z_0$ is a polynomial of $\{\psi_j,\, \overline{\psi}_j\}_{j=1}^n$ with zero free term, there exists $l \in \mathbb{N}$ such that $(\chi - z_0)^l = 0$, and hence the series terminates after a finite number of terms.

The integral over the Grassmann variables is a linear functional, defined on the basis by the relations
\begin{equation*}
\int d\psi_j=\int d\overline{\psi}_k = 0, \qquad \int \psi_j d\psi_j=\int \overline{\psi}_k d\overline{\psi}_k = 1.
\end{equation*}

A multiple integral is defined to be the repeated integral. Moreover ``differentials'' $\{d\psi_j,\, d\overline{\psi}_j\}_{j=1}^n$ anticommute with each other and with $\{\psi_j,\, \overline{\psi}_j\}_{j=1}^n$. Hence for a function~$f$
\begin{equation*}
f(\psi_1, \ldots, \psi_n) = a_0 + \sum\limits_{j=1}^n a_j \psi_j + \ldots + a_{1, \ldots, n} \prod\limits_{j=1}^n \psi_j
\end{equation*}
we have by definition
\begin{equation*}
\int f(\psi_1, \ldots, \psi_n)d\psi_n \ldots d\psi_1 = a_{1, \ldots, n}.
\end{equation*}

The use of Grassmann variables for computing averages of determinants rests on the following identity, valid for any $n \times n$ matrix $A$:
\begin{equation}
\label{gauss_int}
\int \exp\bigg\{ -\sum\limits_{j,k=1}^n \overline{\psi}_jA_{jk}\psi_k \bigg\} \prod\limits_{j=1}^n d\overline{\psi}_j d\psi_j = \det{A}.
\end{equation}
The one more identity is the Hubbard-Stratonovich transformation

\begin{equation}
\label{sqrt_of_exponent}
\begin{gathered}
e^{y^2} = \frac{a}{\sqrt{\pi}}\int e^{2axy - a^2x^2} dx,\\
e^{yt} = \frac{a^2}{\pi}\int e^{ay(u + iv) + at(u - iv) - a^2u^2 - a^2v^2} du dv.
\end{gathered}
\end{equation}
which valid for any complex numbers $y$, $t$ and any positive number $a$. The identities \eqref{sqrt_of_exponent} also hold when $y$, $t$ are arbitrary even Grassmann variables (i.e.\ sums of the products of even number of Grassmann variables). For even Grassmann variables the formulas can be proved by Taylor-expanding $e^{2axy}$ and $e^{ay(u + iv) + at(u - iv)}$ into the series and integrating each term.


The properties explained so far suffice to obtain the integral representation for $F_{2m}$, $m = 1$, whereas the general case $m > 1$ requires some additional preliminaries, pertaining to antisymmetric tensor products. Further details about antisymmetric tensor products may be found in \cite[Chapter 8.4]{Vi:03}.

\subsubsection{Grassmann variables and the exterior product}

The exterior product of vectors is well-known, as well as the exterior product of alternating multilinear forms (see \cite{Vi:03}). However, to prove Proposition \ref{pr1(integral representation)} we need the exterior product of alternating operators. Define 
it as following. Let $A$ be a linear operator on $\Lambda^q \mathbb{C}^n$ 
and $B$ be a linear operator on $\Lambda^r \mathbb{C}^n$. Then the exterior product $A \wedge B$ is the restriction of the linear operator $\operatorname{Alt} \circ (A \otimes B)$ on the $\Lambda^{q + r} \mathbb{C}^n$. Here $\operatorname{Alt}$ is the operator of the alternation, i.e.,
\begin{equation*}
\operatorname{Alt}(t) = \frac{1}{k!}\sum\limits_{\pi \in S_k} \operatorname{sgn} \pi f_\pi (t), \quad t \in \Lambda^k V,
\end{equation*}
where $S_k$ is the group of permutations of length $k$; $\operatorname{sgn} \pi$ is the sign of permutation $\pi$; $f_\pi$ is the canonical automorphism of $V^{\otimes k}$, which carries $v_1 \otimes \ldots \otimes v_k$ to $v_{\pi(1)} \otimes \ldots \otimes v_{\pi(k)}$, $v_j \in V$; $V$ is some finite-dimensional linear space. Note, that for $A \in \End V$ the exterior product $A \wedge A$ coincides with the well-known second exterior power of linear operator $A$.


Fix some basis $\{ e_j \}_{j = 1}^n$ of $\mathbb{C}^n$. Let $A \in \End\Lambda^k \mathbb{C}^n$ and $\alpha, \beta \in I_{n,k}$, where $I_{n, k}$ is defined in \eqref{eq:I_nk definition}. By $A_{\alpha\beta}$ we denote the corresponding entry of the matrix of $A$ in the basis $\{e_{\alpha_1} \wedge \ldots \wedge e_{\alpha_k},\, \alpha \in I_{n,k}\}$.

To obtain the integral representation for $F_{2m}$ we use the lemma:

\begin{lemma}
\label{lemma_1}
Let $A$ and $B$ be linear operators on $\Lambda^q \mathbb{C}^n$ and $\Lambda^r \mathbb{C}^n$ respectively. Then
\begin{multline*}
\sum\limits_{\alpha, \beta \in I_{n, q}} A_{\alpha\beta} \prod\limits_{j = 1}^q \overline{\psi}_{\alpha_j}\psi_{\beta_j} \cdot \sum\limits_{\gamma, \delta \in I_{n, r}} B_{\gamma\delta} \prod\limits_{j = 1}^r \overline{\psi}_{\gamma_j}\psi_{\delta_j} \\
= \binom{q + r}{q} \sum\limits_{\alpha, \beta \in I_{n, q + r}}(A \wedge B)_{\alpha\beta} \prod\limits_{j = 1}^{q + r} \overline{\psi}_{\alpha_j}\psi_{\beta_j}.
\end{multline*}
\end{lemma}

\emph{Proof.} Let $S_{q, r}$ be the set of such $\pi \in S_{q + r}$ that satisfy inequalities $\pi(1) < \ldots < \pi(q)$ and $\pi(q + 1) < \ldots < \pi(q + r)$. Then
\begin{multline*}
\sum\limits_{\alpha, \beta \in I_{n, q}} A_{\alpha\beta} \prod\limits_{j = 1}^q \overline{\psi}_{\alpha_j}\psi_{\beta_j} \cdot \sum\limits_{\gamma, \delta \in I_{n, r}} B_{\gamma\delta} \prod\limits_{j = 1}^r \overline{\psi}_{\gamma_j}\psi_{\delta_j} \\
= \sum\limits_{\alpha, \beta \in I_{n, q + r}}\sum\limits_{\pi, \sigma \in S_{q, r}} \operatorname{sgn} \pi \operatorname{sgn} \sigma A_{\alpha_{\pi}' \beta_{\sigma}'} B_{\alpha_{\pi}'' \beta_{\sigma}''} \prod\limits_{j = 1}^{q + r} \overline{\psi}_{\alpha_j}\psi_{\beta_j},
\end{multline*}
where
\begin{align*}
\alpha_{\pi} &= (\alpha_{\pi(1)}, \ldots, \alpha_{\pi(q + r)}), \\
\alpha' &= (\alpha_{1}, \ldots, \alpha_{q}) \in I_{n,q}, \\
\alpha'' &= (\alpha_{q + 1}, \ldots, \alpha_{q + r}) \in I_{n,r}.
\end{align*}
On the other hand,
\begin{multline*}
(A \otimes B)(e_{\beta'} \wedge e_{\beta''}) = (A \otimes B)\bigg(\frac{1}{(q + r)!} \sum\limits_{\sigma \in S_{q + r}} \operatorname{sgn} \sigma f_\sigma (e_{\beta'} \otimes e_{\beta''})\bigg) \\
= \frac{q!r!}{(q + r)!} (A \otimes B)\bigg( \sum\limits_{\sigma \in S_{q,r}} \operatorname{sgn} \sigma f_\sigma (e_{\beta'} \otimes e_{\beta''})\bigg) \\
= \frac{q!r!}{(q + r)!} \sum\limits_{\substack{\alpha \in I_{n,q} \\ \gamma \in I_{n,r}}} \bigg( \sum\limits_{\sigma \in S_{q, r}} \operatorname{sgn} \sigma A_{\alpha\beta_\sigma'} B_{\gamma\beta_\sigma''}\bigg) e_{\alpha} \otimes e_{\gamma},
\end{multline*}
where $e_\alpha = e_{\alpha_1} \wedge \ldots \wedge e_{\alpha_q}$, $\alpha \in I_{n, q}$.

Hence, 
\begin{multline*}
\operatorname{Alt}((A \otimes B)(e_{\beta'} \wedge e_{\beta''})) = \frac{q!r!}{(q + r)!} \sum\limits_{\substack{\alpha \in I_{n,q} \\ \gamma \in I_{n,r}}} \bigg( \sum\limits_{\sigma \in S_{q, r}} \operatorname{sgn} \sigma A_{\alpha\beta_\sigma'} B_{\gamma\beta_\sigma''}\bigg) e_{\alpha} \wedge e_{\gamma} \\
= \frac{q!r!}{(q + r)!} \sum\limits_{\alpha \in I_{n,q + r}} \sum\limits_{\pi, \sigma \in S_{q, r}} \operatorname{sgn} \sigma A_{\alpha_\pi' \beta_\sigma'} B_{\alpha_\pi'' \beta_\sigma''} e_{\alpha_\pi'} \wedge e_{\alpha_\pi''} \\
= \frac{q!r!}{(q + r)!} \sum\limits_{\alpha \in I_{n,q + r}} \sum\limits_{\pi, \sigma \in S_{q, r}} \operatorname{sgn} \pi \operatorname{sgn} \sigma A_{\alpha_\pi' \beta_\sigma'} B_{\alpha_\pi'' \beta_\sigma''} e_{\alpha'} \wedge e_{\alpha''}.
\end{multline*}

Thus, 
\begin{equation*}
(A \wedge B)_{\alpha\beta} =  \frac{q!r!}{(q + r)!} \sum\limits_{\pi, \sigma \in S_{q, r}} \operatorname{sgn} \pi \operatorname{sgn} \sigma A_{\alpha_\pi' \beta_\sigma'} B_{\alpha_\pi'' \beta_\sigma''},
\end{equation*}
which completes the proof of the lemma. $\blacksquare$

We also need some properties of the exterior product of the operators.

\begin{proposition}
\label{proposition_1}
Let $A_j \in \operatorname{End} \Lambda^{q_j} \mathbb{C}^n$, $j = \overline{1, k}$, and $B \in \operatorname{End} \mathbb{C}^n$. Then
\begin{itemize}

\item[(i)] $A_1 \wedge A_2 = A_2 \wedge A_1$;

\item[(ii)] $(A_1 \wedge A_2) \wedge A_3 = A_1 \wedge (A_2 \wedge A_3)$;

\item[(iii)] $\bigwedge\limits_{j = 1}^k A_j = \Big( \operatorname{Alt} \circ \bigotimes\limits_{j = 1}^k A_j \Big)\Big|_{\Lambda^q \mathbb{C}^n}$;

\item[(iv)] $\bigwedge\limits_{j = 1}^k A_jB^{\wedge q_j} = \Big( \bigwedge\limits_{j = 1}^k A_j \Big)B^{\wedge q}$ and $\bigwedge\limits_{j = 1}^k B^{\wedge q_j}A_j = B^{\wedge q}\Big( \bigwedge\limits_{j = 1}^k A_j \Big)$;
\end{itemize}
where $q = \sum\limits_{j = 1}^k q_j$, $B^{\wedge q} = \underbrace{B \wedge \ldots \wedge B}_{q \text{ times}}$.
\end{proposition}

\emph{Proof.} Assertions (i) and (ii) follow from Lemma 1 and from Grassmann variables multiplication's anticommutativity and associativity.

(iii) Consider the case $k = 3$.
\begin{multline*}
A_1 \wedge A_2 \wedge A_3 = \operatorname{Alt} \circ ((I \circ A_1) \otimes (\operatorname{Alt} \circ (A_2 \otimes A_3))) \\
= \operatorname{Alt} \circ (I \otimes \operatorname{Alt}) \circ (A_1 \otimes A_2 \otimes A_3) = \operatorname{Alt} \circ (A_1 \otimes A_2 \otimes A_3),
\end{multline*}
where $I$ is the identity operator.

The general case follows from the induction.

(iv) By definition, we have
\begin{equation*}
\bigwedge\limits_{j = 1}^k B^{\wedge q_j}A_j = \operatorname{Alt} \circ \bigotimes\limits_{j = 1}^k B^{\wedge q_j} A_j.
\end{equation*}
Consider $(\operatorname{Alt} \circ B^{\otimes q_j})(v_1 \otimes \ldots \otimes v_{q_j})$, $v_l \in \mathbb{C}^n$
\begin{multline*}
(\operatorname{Alt} \circ B^{\otimes q_j})(v_1 \otimes \ldots \otimes v_{q_j}) = \frac{1}{q_j!} \sum\limits_{\pi \in S_{q_j}} \operatorname{sgn} \pi f_\pi (Bv_1 \otimes \ldots \otimes Bv_{q_j}) \\
= \frac{1}{q_j!} \sum\limits_{\pi \in S_{q_j}} \operatorname{sgn} \pi Bv_{\pi(1)} \otimes \ldots \otimes Bv_{\pi(q_j)} \\
= \frac{1}{q_j!} \sum\limits_{\pi \in S_{q_j}} \operatorname{sgn} \pi B^{\otimes q_j}(f_\pi (v_1 \otimes \ldots \otimes v_{q_j})) = (B^{\otimes q_j} \circ \operatorname{Alt})(v_1 \otimes \ldots \otimes v_{q_j}).
\end{multline*}
Therefore, $\operatorname{Alt} \circ B^{\otimes q_j} = B^{\otimes q_j} \circ \operatorname{Alt}$, in particular, $B^{\wedge q_j} = \left. B^{\otimes q_j} \right|_{\Lambda^{q_j} \mathbb{C}^n}$. Thus, 
\begin{align*}
\bigwedge\limits_{j = 1}^k B^{\wedge q_j}A_j = \operatorname{Alt} \circ \bigotimes\limits_{j = 1}^k B^{\otimes q_j} A_j &= \operatorname{Alt} \circ B^{\otimes q} \circ \bigg(\bigotimes\limits_{j = 1}^k A_j\bigg) \\
&= B^{\otimes q} \circ \operatorname{Alt} \circ \bigg(\bigotimes\limits_{j = 1}^k A_j\bigg) = B^{\wedge q}\bigg( \bigwedge\limits_{j = 1}^k A_j \bigg).
\end{align*}

The proof of the second formula is similar. $\blacksquare$

\subsection{The Harish-Chandra/Itsykson--Zuber formula}

For computing the integral over the unitary group, the following Harish-Chan\-dra/It\-syk\-son--Zuber formula is used
\begin{proposition}\label{pr:H-C/I--Z formula}
Let $A$ be a normal $n \times n$ matrix with distinct eigenvalues $\{a_j\}_{j = 1}^n$ and $B = \diag\{b_j\}_{j = 1}^n$, $b_j \in \mathbb{R}$. Then
\begin{equation*}
\int\limits_{U_n} \exp\{z\tr AU^*BU\}dU_n(U) = \bigg(\prod\limits_{j = 1}^{n - 1} j!\bigg) \frac{\det\{\exp(za_jb_k)\}_{j,k = 1}^n}{z^{(n^2 - n)/2}\Delta(A)\Delta(B)},
\end{equation*}
where $z$ is come constant, $\Delta(A) = \Delta(\{a_j\}_{j = 1}^n)$, $\Delta$ is defined in \eqref{eq:Vandermonde definition}. Moreover, for any symmetric domain $\Omega$ and any symmetric function $f(B)$ of $\{b_j\}_{j = 1}^n$
\begin{multline}\label{eq:H-C/I--Z (symm.domain)}
\int\limits_{U_n}\int\limits_{\Omega} \exp\{z\tr AU^*BU\}\Delta^2(B)f(B)dU_n(U)dB \\
= \bigg(\prod\limits_{j = 1}^n j!\bigg) \frac{z^{-(n^2 - n)/2}}{\Delta(A)}\int\limits_{\Omega} \exp\bigg\{z\sum\limits_{j = 1}^n a_jb_j\bigg\}\Delta(B)f(B) dB,
\end{multline}
where $dB = \prod\limits_{j = 1}^{n} db_j$.
\end{proposition}
For the proof see, e.g., \cite[Appendix 5]{Me:91}.

\begin{remark}
Notice, that \eqref{eq:H-C/I--Z (symm.domain)} is also valid if $A$ has equal eigenvalues.
\end{remark}
Indeed, if $a_{j_1} = a_{j_2}$, $j_1 \ne j_2$, the integrand in \eqref{eq:H-C/I--Z (symm.domain)} changes sign when $b_{j_1}$ and $b_{j_2}$ are swapped. Thus, the ratio of the integral and $a_{j_1} - a_{j_2}$ is well defined.

\section*{Acknowledgement}
The author is grateful to Prof.\ M.\ Shcherbina for statement of the problem and for fruitful discussions. The author is thankful to the Akhiezer Foundation for partial financial support.

\newcommand{\vol}[1]{\textbf{#1}}
\bibliographystyle{spmpsci}

\end{document}